\documentclass{memoir}
\usepackage[english]{babel}
\usepackage[T1]{fontenc}
\usepackage[utf8]{inputenc}
\usepackage[numbers,sort&compress]{natbib}
\usepackage{listings}
\usepackage{graphicx}
\usepackage{changepage}
\usepackage{lstautogobble}
\usepackage[hidelinks]{hyperref}
\usepackage{helvet}
\usepackage{siunitx}

\usepackage{tikz, pgfplots}
\pgfplotsset{compat=1.18}
\usetikzlibrary{positioning, shapes.multipart, shapes.geometric, automata}
\usepackage[%
    left=1in,%
    right=1in,%
    top=1.0in,%
    bottom=1.0in,%
    paperheight=11in,%
    paperwidth=8.5in%
]{geometry}%

\definecolor{darkred}{HTML}{d62728}
\usepackage{amsmath}
\counterwithout{figure}{chapter}
\counterwithout{figure}{section}
\counterwithout{equation}{section}
\counterwithout{equation}{chapter}
\counterwithout{table}{chapter}
\counterwithout{table}{section}
\begin{document}

\date{}



\chapter*{Performing all-atom molecular dynamics simulations of intrinsically disordered proteins with replica exchange solute tempering}


\chapterprecis{Jaya Krishna Koneru, Korey M. Reid and Paul Robustelli}

Dartmouth College, Department of Chemistry, Hanover, NH, 03755, USA

\section*{Abstract}

All--atom molecular dynamics (MD) computer simulations are a valuable tool for characterizing the conformational ensembles of intrinsically disordered proteins (IDPs). 
IDP conformational ensembles are highly heterogeneous and contain structures with many distinct topologies separated by large free--energy barriers. 
Sampling the vast conformational space of IDPs in explicit solvent all--atom MD simulations is extremely challenging, and enhanced sampling methods are generally required to obtain statistically meaningful descriptions of IDP conformational ensembles. 
Replica exchange solute tempering (REST) methods, where multiple coupled simulations of a system are performed in parallel with selectively modified potential energy functions, are a powerful approach for efficiently sampling the conformational space of IDPs. In this chapter, we demonstrate how to set--up, perform and analyze all--atom MD simulations of IDPs with REST enhanced sampling methods. 

\vspace{5mm}

\textbf{Keywords} Intrinsically disordered proteins, molecular dynamics, replica exchange, solute tempering, simulation convergence   

\section*{Introduction}\label{sec:Intro}

 
Many proteins that perform important biological function do not adopt stable tertiary structures in isolation under physiological conditions~\cite{holehouse2024molecular,banerjee2023dissecting}. 
These intrinsically disordered proteins (IDPs) populate a heterogeneous conformational ensemble of rapidly interconverting structures in solution. 
Experimentally characterizing the conformational ensembles of IDPs in atomic detail is extremely challenging, as most biophysical experimental techniques used to structurally characterize IDPs in solution report on conformational properties averaged over many molecules over long periods of time~\cite{bonomi2017principles}. 
All--atom molecular dynamics (MD) computer simulations are a powerful approach for obtaining atomic resolution descriptions of IDP conformational ensembles and can provide mechanistic molecular insight into their biological interactions and functions~\cite{Robustelli2018,sisk2024folding,zhu2022small}. 
Recent improvements in the physical models, or \textit{force fields}, used in MD simulations have dramatically improved the accuracy of IDP simulations~\cite{Robustelli2018,Piana2015,huang2017charmm36m,best2014balanced,Piana2020}. 

The conformational ensembles of IDPs are extremely heterogeneous and contain a large number of structurally distinct conformations separated by large free--energy barriers. 
As a result, sampling the conformational space of IDPs in MD simulations is notoriously challenging. 
Conventional unbiased \textit{continuous-time} MD simulations, which simulate the structural evolution of a single IDP system in time, are highly likely to become trapped in local free energy minima and may only sample a small fraction of the accessible conformational space of an IDP in currently feasible simulation times. 
Enhanced sampling methods~\citep{henin2022enhanced} are generally required to meaningfully sample the conformational space of IDPs. 

Replica exchange methods~\citep{Sugita1999,hansmann1997parallel} are a powerful approach for sampling the conformational ensembles of IDPs in all--atom MD simulations. 
In these approaches, multiple copies (or \textit{replicas}), of a system are simulated in parallel at different temperatures or with selectively modified energy functions (or \textit{Hamiltonians}), and Monte Carlo exchange attempts are used to stochastically exchange coordinates between replicas at regular intervals. In temperature replica exchange MD (tREMD) each replica of a system is simulated with the same potential energy function at a different temperature. 
In tREMD simulations, replicas are sometimes referred to as \textit{temperature rungs} spanning a so--called \textit{temperature ladder}. 
Replicas simulated at high temperatures can efficiently overcome free--energy barriers and replicas simulated at lower temperatures rigorously sample the unbiased Boltzmann--weighted conformational distribution at physiologically relevant temperatures of interest. 
A practical challenge for applying tREMD to explicit solvent biomolecular simulations is that the number of replicas required to obtain efficient exchanges across a given temperature ladder scales exponentially with the degrees of freedom of the simulated system. Depending on the size of the system, simulating a protein in explicit water box on a temperature ladder spanning a few hundred degrees Kelvin (ie. 300K--400K or 300K--600K) may require tens to hundreds of replicas. 
This makes explicit solvent tREMD simulations of large biomolecules impractical without access to massively parallel computing resources. 

Replica exchange with solute tempering (REST) methods provide an elegant solution to this problem~\citep{Liu2005}. 
In the REST approach, the Hamiltonian of each temperature rung in a replica exchange simulation is modified such that only the solute (non--water) atoms of the system are heated. This dramatically reduces the number of replicas required to enable efficient Monte Carlo exchanges across a given temperature ladder. 
A more efficient variant of the REST method, referred to as REST2, and has become widely adopted in bimolecular simulation community~\citep{Wang2011}. This has been facilitated by a flexible implementation of REST methods~\citep{bussi2014hamiltonian} in the popular GROMACS simulation engine~\citep{VanDerSpoel2005,abraham2015gromacs} using the open--source PLUMED enhanced sampling plug--in~\citep{Bonomi2009,Tribello2014}. 
A number of variants of the REST method, which explore a variety of modified hamiltonian scaling schemes have been proposed~\citep{kamiya2018flexible,lao2024replica}. Recent studies have shown that IDPs can collapse into compact conformations at high solute temperatures in REST2 simulations, potentially frustrating sampling, and two recently developed variants of the REST method appear to have particular advantages for simulating IDPs~\citep{Appadurai2021,Zhang2023}. In the replica exchange with hybrid tempering method (REHT)~\citep{Appadurai2021}, the REST2 scheme is modified such that solvent molecules are heated alongside solute molecules using a separate more gradual temperature ladder~\cite{Appadurai2021}.
In the REST3 method~\citep{Zhang2023}, an additional Hamiltonian scaling factor is introduced to scale solute--solvent interactions to prevent the collapse of IDP conformations at high solute temperatures. 

In this chapter, we provide an overview of the theory of REST methods and a step--by--step guide on how to set--up, perform, and analyze an all--atom REST2 MD simulation of an IDP in the GROMACS simulation engine with the PLUMED enhanced sampling plug--in.
As an example, we demonstrate how to perform a REST2 simulation of a previously studied 20--residue disordered fragment of the IDP $\alpha$--synuclein~\citep{Robustelli2022}. 
All GROMACS simulation input files, starting structures, and analysis code for this example are freely provided in the accompanying GitHub repository \url{https://github.com/paulrobustelli/IDP_REST_tutorial}.



 

\subsection*{\textit{Replica Exchange with Solute Tempering (REST) Theory}}\label{sec:Intro-RE}
In classical molecular dynamics (MD) simulations, the evolution of the coordinates of a molecular system in time is dictated by an energy function known as a Hamiltonian. 
The Hamiltonian, which describes both the kinetic energy and potential energy of a molecular system, is used to compute the forces between all atoms at each timestep in a simulation, and the calculated forces are subsequently used to update the positions of the atoms. 
Conventional MD simulations provide a continuous description of how the positions of atoms in a simulated molecular system fluctuate in time and can be used to calculate dynamical properties, such as rates of transitions between conformational states or autocorrelation functions of properties of interest. 

In temperature replica exchange molecular dynamics (tREMD) simulations, multiple non--interacting copies, or \textit{replicas}, of a molecular system are simulated in parallel at a set of $N$ different temperatures $\{T_{0},T_{1},T_{2},...,T_{N}\}$, referred to as a \textit{temperature ladder}~\citep{Sugita1999}. 
After simultaneously performing a specified number of simulation timesteps of each replica in parallel, attempts are made to exchange the atomic coordinates of pairs of neighboring temperature replicas. 
The proposed coordinate exchanges, or \textit{swaps}, are accepted or rejected according to a Metropolis acceptance criteria based on the potential energy of simulated system at each temperature. 
The metropolis acceptance criteria, which is based on the probability that a set of coordinates with a given potential energy sampled at one temperature would be sampled in a different simulation of the same system at a different temperature, preserves the condition of detailed balance among all replicas in the simulation, and ensures the correct statistical sampling of the Boltzmann distribution at each temperature.  
This process is repeated at regular intervals until the tREMD simulation is ended. 

In tREMD simulations, "hot" replicas simulated at higher temperatures should cross conformational free--energy barriers more frequently, accelerating the sampling of conformational space. 
Simulated configurations from hot replicas then diffuse throughout the temperature ladder via Metropolis exchange attempts and the configurations sampled by at each temperature rung rigorously correspond to the unbiased Boltzmann--weighted canonical ensemble of that temperature. In the ergodic limit of infinite sampling, the probability distribution of conformations sampled in each temperature rung in tREMD and the probability distribution of conformations sampled in conventional continuous--time MD simulations run at the same temperature should converge to the same equilibrium probability distribution. 
Performing tREMD with an appropriate temperature ladder can dramatically accelerate the sampling of conformational processes of interest in biomolecular systems. 
We note that by using tREMD with frequent exchange attempts, information about the continuous--time dynamics of the system at each temperature is largely lost. 

In conventional tREMD simulations, the same Hamiltonian is used for all replicas and the temperature of each replica is controlled using a simulation thermostat~\citep{Bussi2007}. 
In tREMD each temperature rung corresponds to a physically meaningful simulation temperature. 
In contrast, replica exchange with solute tempering (REST) methods can be considered as a subclass of Hamiltonian replica exchange molecular dynamics (HREMD) methods. 
In these approaches the Hamiltonian of each replica can be modified in arbitrary manner to accelerate sampling of specific conformational changes of interest.  The probability distributions obtained in replicas with selectively scaled Hamiltonians may, however, no longer correspond to a realistic physical temperature or system.

In the original REST implementation for protein simulations~\citep{Liu2005}, the Hamiltonian of each of the $n$ replicas ($E_{n}^{REST}$) was decomposed into three contributions corresponding to the internal energy of the protein ($E_{pp}$), the interaction energy between the protein and water ($E_{pw}$), and the interaction energy of water molecules with other water molecules ($E_{ww}$):
\begin{center}
    \begin{equation}
        E_{n}^{REST}(X_{n}) = E_{pp}(X_{n}) + E_{pw}(X_{n}) + E_{ww} (X_{n})
    \label{eq:remd_hamiltonian}
    \end{equation}
\end{center}
where $X_{n}$ represents a molecular configuration of a simulated system from the $n^{th}$ replica. 
The Hamiltonian of each replica is subsequently modified by introducing scaling factors for each of these terms
\begin{center}
    \begin{equation}
        E_{n}^{REST}(X_{n}) = \lambda_{n}^{pp} E_{pp}(X_{n}) + \lambda_{n}^{pw} E_{pw}(X_{n}) + \lambda_{n}^{ww} E_{ww} (X_{n})
    \label{eq:hremd_hamiltonian}
    \end{equation}
\end{center}
where $\lambda_{n}^{pp}$, $\lambda_{n}^{pw}$ and $\lambda_{n}^{ww}$ are the scaling factors for protein--protein interactions, protein--water interactions, and water--water interactions, respectively, for the $n^{th}$ replica. 
In the initial REST implementation~\citep{Liu2005} (REST1), the scaling factors for each replica are determined according to
\begin{center}
    \begin{equation}
        E_{n}^{REST1}(X_{n}) =  E_{pp}(X_{n}) + \frac{\beta_{0}+\beta{n}}{2\beta_{n}} E_{pw}(X_{n}) +  \frac{\beta_{0}}{\beta_{n}}E_{ww} (X_{n})
        \label{eq:rest1_hamiltonian}
    \end{equation}
\end{center}
\begin{center}
    \begin{equation} 
        \beta_{n}=\frac{1}{k_{B}T_{n}}
    \label{eq:REST1_lambda_scaling}     
    \end{equation}
\end{center}
where $T_{n}$ is the selected temperature of replica $n$.  
In REST1, each replica is simulated at a different temperature ($T_{n}$), and with a different potential energy as specified in Eq.~\ref{eq:rest1_hamiltonian}. 

It was subsequently demonstrated that REST1 could be less efficient than tREMD for larger biomolecular systems systems~\citep{huang2007replica}. 
A new implementation of REST, referred to as REST2, was subsequently proposed to address this deficiency~\cite{Wang2011}. 
In the initial REST2 implementation, every replica is simulated at the same temperature (ie. the thermostat of each replica is set to the same value of $T_{0}$), and the Hamlitonion of the system is selectively scaled for each replica with scaling factors according to:
\begin{center}
    \begin{equation}
            E_{n}^{REST2}(X_{n}) = \frac{\beta_{n}}{\beta_{0}}E_{pp}(X_{n}) + \sqrt{\frac{\beta_{n}}{\beta_{0}}} E_{pw}(X_{n}) + E_{ww} (X_{n})
        \label{eq:rest2_hamiltonian}
    \end{equation}
\end{center}
In REST2, while the thermostat of each replica is set to the same value of $T_{0}$, one can consider the \textit{effective solute temperature} ($T_{solute}$) of the modified Hamiltonian based on the scaling factor applied to $E_{PP}$ according to: 

\begin{center}
    \begin{equation}
 \frac{\beta_{n}}{\beta_{0}}=\frac{T_{0}}{T_{n}}.  
    \label{eq:SoluteTemp}
    \end{equation}
\end{center}
For example, if $T_{0}$ = 300 K a scaling factor of $\frac{\beta_{n}}{\beta_{0}}=0.5$ corresponds to an effective solute temperature of $T_{solute}$ = 600 K. 
That is, scaling the potential energy function of a solute by a factor of 0.5 and simulating with a thermostat set to $T$ = 300 K is effectively equivalent to using the original unscaled potential energy function and employing an independent solute thermostat with a temperature of $T$ = 600 K.

Scaling factors are often determined using a solute temperature ladder that follows a logarithmic scale or geometric scale~\citep{prakash2011replica}. 
A solute temperature ladder with logarithmic scale can be defined as:
\begin{equation}
    T_n=T_0*\exp{\Bigg[\frac{n*log(T_{high}/T_0)}{N_r}\Bigg]}
\end{equation}
Where $T_0$ is the solute temperature of the lowest temperature replica and $T_{high}$ is the solute temperature of the highest rung of temperature ladder. 
For each desired solute temperature, the scaling factors in Eq.~\ref{eq:rest2_hamiltonian} are set according to Eq.~\ref{eq:SoluteTemp} After the desired solute temperature ladder has been specified and the appropriate REST2 scaling factors have been used to specify the modified Hamiltonian of each replica, the acceptance criteria of a proposed exchange of coordinates sampled in a replica from solute temperature rung $n$ ($X_{n}$) and the coordinates from a neighboring solute temperature rung $n+1$ ($X_{n+1}$) is given by:
\begin{center}
    \begin{equation}
        \begin{aligned}
        \Delta_{n,n+1} = (\beta_{n} - \beta_{n+1})& \Big[(E_{pp}(X_{n+1}) - E_{pp}(X_{n})) \\
        &+ \frac{\sqrt{\beta_{0}}} {\sqrt{\beta_{n}}+\sqrt{\beta_{n+1}}} (E_{pw}(X_{n+1}) - E_{pw}(X_{n}))\Big].
        \end{aligned}
        \label{eq:rest2_AR}
    \end{equation}
\end{center}

If $\Delta_{n,n+1}$ is negative, the configuration ($X_{n+1}$) is more energetically favorable than then configuration ($X_{n}$) and the proposed swap is accepted. If $\Delta_{n,n+1}$ is positive, the configuration ($X_{n+1}$) is less energetically favorable than then configuration ($X_{n}$), and the proposed swap has an acceptance probability ($P_{acc}$) determined according to:

\begin{center}
    \begin{equation}
        \begin{aligned}
     P_\mathrm{acc} = \min(1, \exp(-\Delta_{n,n+1}))
        \end{aligned}
        \label{eq:rest2_acc}
    \end{equation}
\end{center}

We note that the energy of the interactions between water molecules ($E_{ww}$) does not appear in Eq.~\ref{eq:rest2_AR}. 
This means that the energetics of purely solvent degrees of freedom do not affect the exchange probabilities between neighboring replicas in REST2 simulations. 
This results in higher acceptance ratios between solute temperature rungs in REST2 simulations compared to temperature rungs in tREMD simulations, and enables the use of a dramatically smaller number of replicas to span a desired temperature range in REST2 simulations.

In REST2 simulations of disordered proteins, it has been observed that IDPs simulated at high solute temperatures sample highly compact conformations compared to unscaled MD simulations run at the same temperature~\citep{Zhang2023}. 
This over--collapse at high solute temperatures may decrease the sampling efficiency of REST2 simulations of IDPs. 
To remedy this effect a new REST scheme, termed REST3, was proposed that introduces an additional scaling factor to scale the van der Waals (vdW) interactions between protein and water atoms at high temperatures~\citep{Zhang2023}.


\section*{Materials}
\subsection*{\textit{Required Software}}

Here, we review the software required to set--up, run and analyze the REST2 simulation example provided in accompanying online GitHub repository (\url{https://github.com/paulrobustelli/IDP_REST_tutorial}). The example REST2 simulation is produced using GROMACS~\citep{VanDerSpoel2005,abraham2015gromacs,pall2020heterogeneous} patched with the PLUMED2 enhanced sampling engine~\citep{Bonomi2009,Tribello2014}. Visual Molecular Dynamics (VMD)~\citep{HUMP96} and PyMOL~\cite{PyMOL,Yuan2017} are employed to visualize simulated ensembles. Analysis of REST2 trajectories are performed with python code and a local installation of python3~\citep{python3} is required with the the matplotlib~\citep{matplotlibdev2024,Hunter2007}, numpy~\citep{harris2020array} and scipy~\citep{2020SciPy} python packages. The structural properties of MD ensembles (dihedral angles, intramolecular contacts, radius of gyration, and other structural descriptors of interest) are analyzed using the MDTraj~\citep{McGibbon2015} and MDAnalysis~\cite{Michaud-Agrawal2011,Gowers2016} python libraries. Statistical error estimates of simulated quantities are obtained using a Flyvbjerg blocking analysis~\citep{Flyvbjerg1989} performed with the pyblock package~\citep{pyblock}. Instructions to prepare a basic installation of PLUMED, GROMACS and the necessary python environment are provided in Note~\ref{notes:installation} and Note~\ref{notes:pythonEnv}.
\begin{itemize}
  \item \textbf{GROMACS (Gr\"{o}ningen Machine for Chemical Simulations)} 
  A versatile and widely--used molecular dynamics simulation software package mainly designed for simulations of proteins, nucleic acids, and lipids~\citep{VanDerSpoel2005,abraham2015gromacs,pall2020heterogeneous}. The accompanying tutorial was performed with GROMACS Version 2022.5~\citep{Gromacs2022.5}.

  \item \textbf{PLUMED (PLUgin for MolEcular Dynamics)} 
  REST simulations are performed using a version of GROMACS patched with the PLUMED enhanced sampling plugin~\citep{Bonomi2009,Tribello2014}. PLUMED extends the functionalities of traditional MD engines to enable a number of enhanced sampling methods, including REST, umbrella sampling, metadynamics and others. The tutorial was performed with PLUMED 2.9.0.
  
  \item \textbf{VMD (Visual Molecular Dynamics)}
  VMD is a molecular visualization program for displaying, animating, and analyzing large biomolecular systems using 3D graphics and built--in scripting~\citep{HUMP96}. VMD version 1.9.4 was used to visualize structural ensembles in the tutorial.
  
  \item \textbf{PyMOL (The PyMOL Molecular Graphics System)} PyMOL is molecular visualization software that also contains the functionality to construct polypeptides and an interface to perform trajectory analysis. PyMOL version 3.0 was used to prepare structures and visualize structural ensembles in this tutorial. 
  
  \item \textbf{Python}~\citep{python3}
  Python scripts were run using python version 3.11 with the following libraries:
    \begin{itemize}
    \item \textbf{mdtraj}
    Version 1.9.7~\citep{McGibbon2015}.
    \item \textbf{pyblock}
    Version 0.6~\citep{pyblock}.
    \item \textbf{numpy}
    Version 1.26.4~\citep{harris2020array}.
    \item \textbf{matplotlib}
    Version 3.9.2~\citep{matplotlibdev2024,Hunter2007}.
    \item \textbf{scipy}
    Version 1.14.1~\citep{2020SciPy}.
    \item \textbf{MDAnalysis}
    Version 2.7.0~\citep{Michaud-Agrawal2011,Gowers2016}.
    \end{itemize}
\end{itemize}
 
\subsection*{\textit{Force Field Parameters and Starting Structures}}

Before running an explicit solvent MD simulation of an IDP, one must identify appropriate protein, water and ion force field parameters. We note that many of the most popular AMBER and CHARMM protein force fields and water model combinations, which give excellent results for folded proteins, do not provide accurate descriptions of IDPs~\citep{Robustelli2018,Huang2018,best2014balanced,piana2015water}. It is therefore important to consult the IDP simulation literature to determine if a given protein force field and water model combination is appropriate for simulating an IDP. The accompanying tutorial uses the a99SB--\textit{disp} protein force field and accompanying a99SB--\textit{disp} water model, which has been found to provide accurate descriptions of many IDPs spanning a range of residual secondary structure populations~\citep{Robustelli2018}. In unbiased MD and REST2 simulations, it is also important to select physically reasonable starting structures. Choosing inappropriate or unphysical starting structures can substantially increase the amount of simulation time required to obtain meaningful statistical ensembles of the desired equilibrium conformational distribution. A discussion of options for generating starting structures of IDPs is included in the Methods section and in Note~\ref{notes:secondary_structure}. Force field field files and starting structures are provided for the accompanying tutorial. 

\section*{Methods}

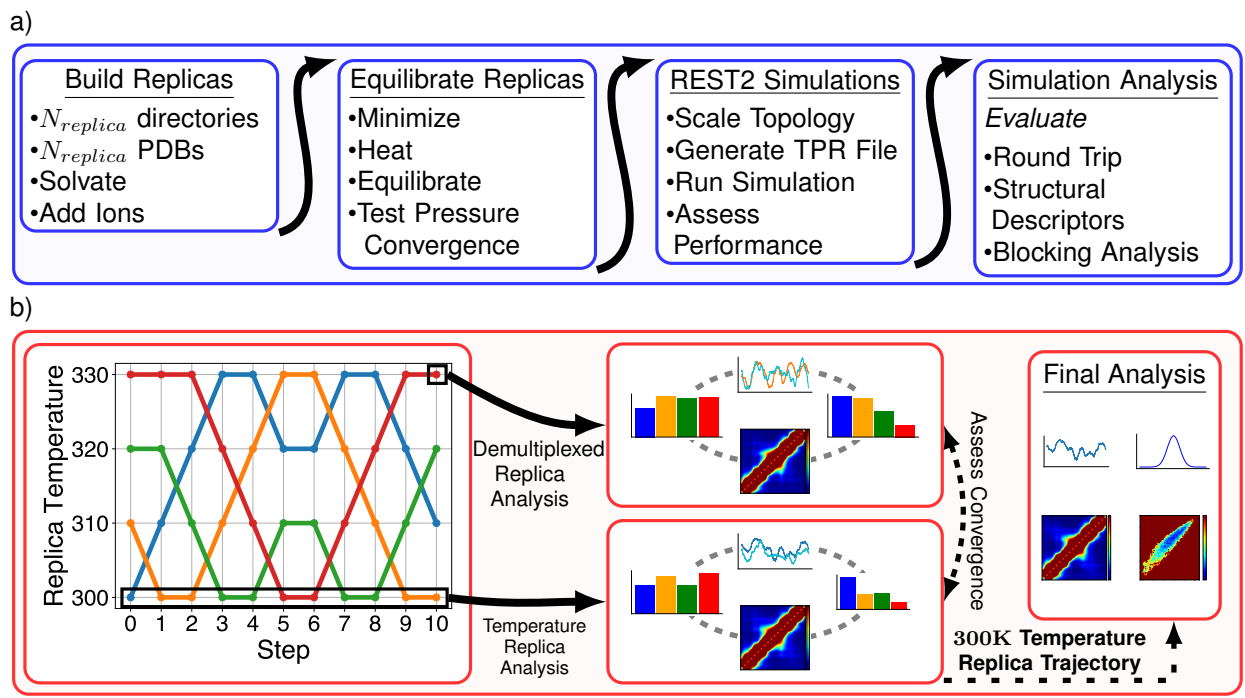
\begin{figure}
    \tikzstyle{arrow} = [thick,->,>=stealth, line width=1.5pt]
    
    \begin{tikzpicture}[
        STEP/.style={rectangle, baseline=(top), text width=1.25in, rounded corners=2mm, draw=blue!80, very thick, minimum size=5mm, fill=white!100},
        INLAY/.style={rectangle, baseline=(top), text width=1.24in, rounded corners=2mm, very thick, minimum size=5mm, anchor=south west, minimum width=6.425in},
        LABEL/.style={circle, inner sep=0pt, radius= 0.02in, draw=none, fill=white, anchor=center, align=left},
        neflat/.style={
            append after command={%
                \pgfextra
                    \fill[fill=#1] (\tikzlastnode.north west)  -- (\tikzlastnode.west) -- (\tikzlastnode.south west)[rounded corners] -| (\tikzlastnode.east)[sharp corners]  -- (\tikzlastnode.north east)  -- cycle;
                    \draw[sharp corners, blue, ultra thick] (\tikzlastnode.south west) -- (\tikzlastnode.south)[rounded corners] -| (\tikzlastnode.north east)  ;
                \endpgfextra}},]
        \node[INLAY, minimum height=1.25in, draw=blue!80, fill=blue!2] at (0,0)  {};
        \node[STEP, anchor=south west] at (0.1,0.7) (Structures) {
                    \centerline{\underline{Build Replicas}}\\
                    [0.5ex]\textbullet $N_{replica}$ directories\\
                    \textbullet $N_{replica}$ PDBs\\
                    \textbullet Solvate\\
                    \textbullet Add Ions
                    };
        \path node[] at (0.05in,1.3725in) (A) {a)}; 

        \node[STEP] (Equilibration) [right=2.5cm of Structures.north, anchor=north west] {\centerline{\underline{Equilibrate Replicas}}\\
                    [0.5ex]\textbullet Minimize\\
                    \textbullet Heat\\
                    \textbullet Equilibrate\\
                    \textbullet Test Pressure \\
                    \hspace{1mm} Convergence};
        
        \node[STEP] (REST2) [right=2.5cm of Equilibration.north, anchor=north west] {\centerline{\underline{REST2 Simulations}}\\
                    [0.5ex]\textbullet Scale Topology\\
                            \textbullet Generate TPR File\\
                            \textbullet Run Simulation\\
                            \textbullet Assess\\\hspace{1mm}Performance
                        };
        
        \node[STEP] (Analysis) [right=2.5 of REST2.north, anchor=north west] {\centerline{\underline{Simulation Analysis}}\\
                    [0.5ex]\textit{Evaluate}\\
                    [0.5ex]\textbullet Round Trip\\
                    \textbullet Structural\\\hspace{1mm}Descriptors\\
                    \textbullet Blocking Analysis};
        
        \draw (Structures.south east) edge[out=15,in=195,-latex, line width=3] (Equilibration.north west);
        \draw (Equilibration.south east) edge[out=15,in=195,-latex, line width=3] (REST2.north west);
        \draw (REST2.south east) edge[out=15,in=195,-latex, line width=3] (Analysis.north west);

    \end{tikzpicture} 
    
    \tikzstyle{boxit} = [rectangle, very thick, minimum height=4pt, draw=black]
    \begin{tikzpicture}[
        STEP/.style={rectangle, baseline=(top), rounded corners=2mm, draw=red!80, very thick, minimum size=5mm, fill=white!100, minimum width=1.75in, anchor=south west},
        COMM/.style={rectangle, text width=1.25in},
        INLAY/.style={rectangle, baseline=(top), text width=1.24in, rounded corners=2mm, very thick, minimum size=5mm, anchor=south west, minimum width=6.425in},
        neflat/.style={
            append after command={%
                \pgfextra
                    \fill[fill=#1] (\tikzlastnode.north west)  -- (\tikzlastnode.west) -- (\tikzlastnode.south west)[rounded corners] -| (\tikzlastnode.east)[sharp corners]  -- (\tikzlastnode.north east)  -- cycle;
                    \draw[sharp corners, red, ultra thick] (\tikzlastnode.south west) -- (\tikzlastnode.south)[rounded corners] -| (\tikzlastnode.north east)  ;
                \endpgfextra}},
        ]
        \node [INLAY, minimum height=1.9in, draw=red!80, fill=red!2] at (0,0)  {};
        \node [STEP] at (0.15,0.15)  (REST2Figure) 
        {
            \includegraphics[scale=0.35]{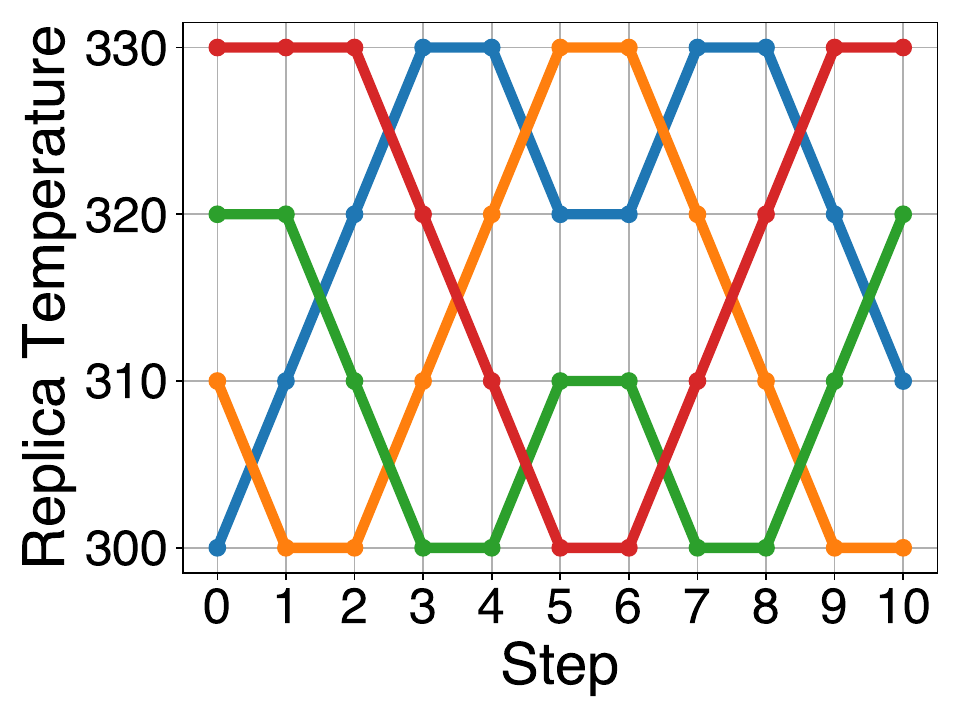}
        };
        \path node[] at (0.05in,2.025in) (B) {b)}; 
        \node [boxit, anchor=south west] at (5.515,4.135) (demuxrep) {};
        \node [boxit, minimum width=123pt, anchor=south west] at (1.44,1.165) (temprep) {};

        \node [STEP, minimum height=0.85in] at (7.9,2.5) (dmfig) {};
        \node
            [draw = gray, dashed, line width=2, ellipse, minimum width = 1in, minimum height = 0.5in, anchor=center]
            (e1) at ([yshift=0.05in]dmfig.center) {};
        \node[anchor=center] at (e1.180)  (dtemp)
                {\includegraphics[width=0.5in]{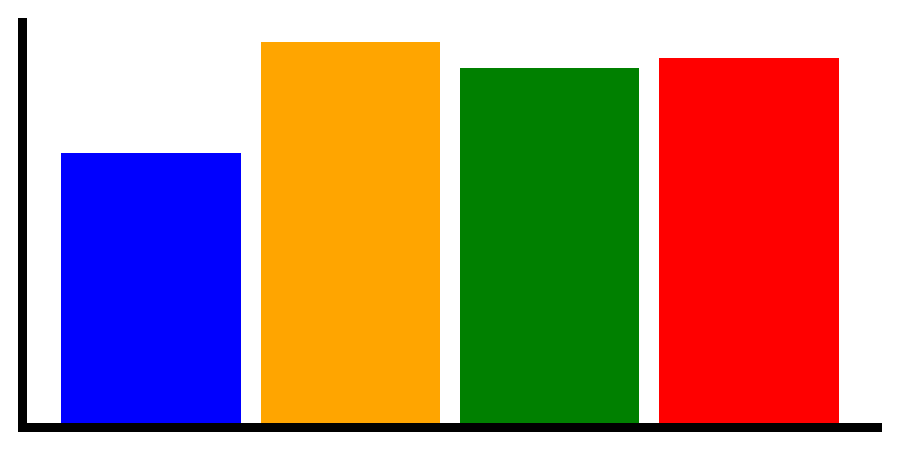}};
        \node[anchor=center] at ([yshift=-0.05in]e1.90)  (drg)
                {\includegraphics[width=0.4in]{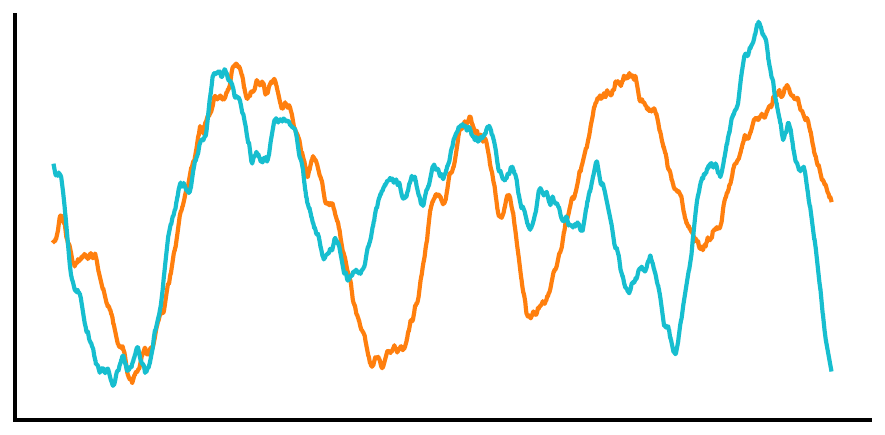}};
        \node[anchor=center] at (e1.0)  (drt)
                {\includegraphics[width=0.5in]{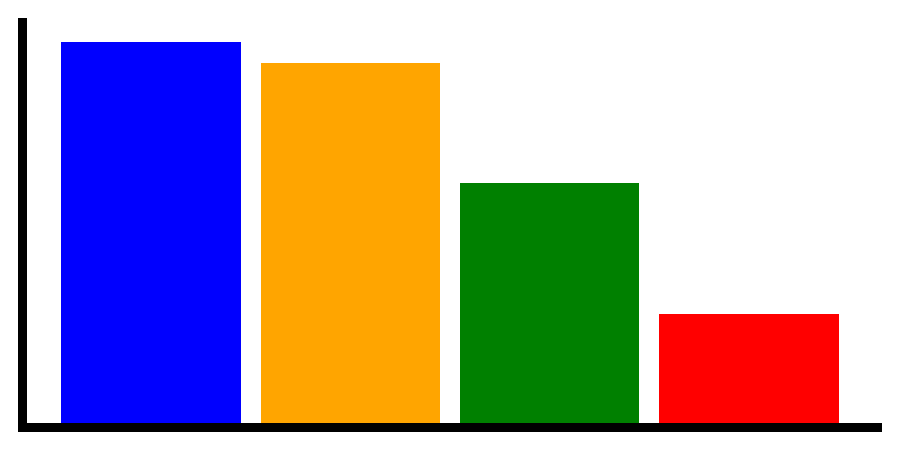}};
        \node[anchor=center] 
                at ([yshift=0.025in]e1.270)  (dcm0) {\includegraphics[height=0.35in]{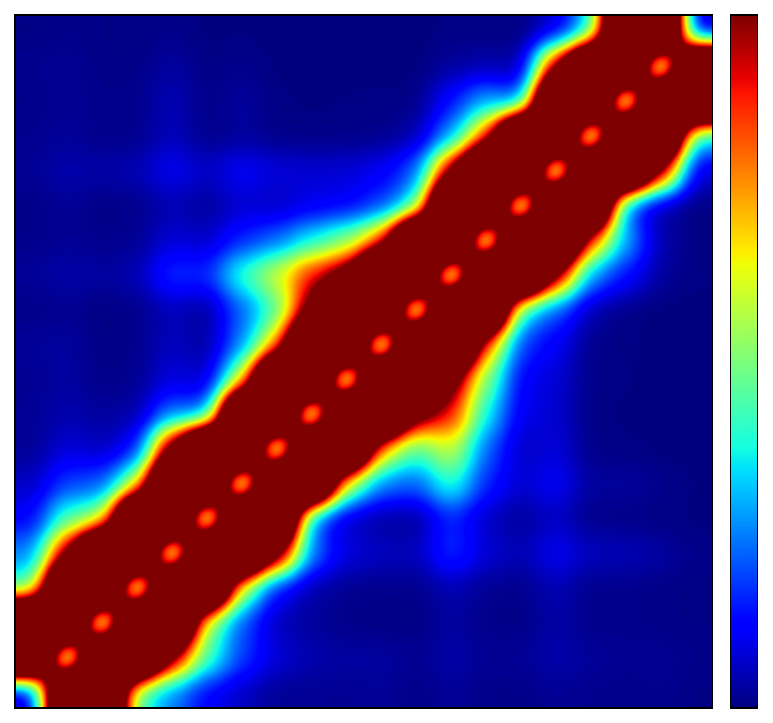}};
        
        \node [STEP, minimum height=0.85in] at (7.9,0.15) (trfig) {};
        \node
                [draw = gray, dashed, line width=2, ellipse, minimum width = 1in, minimum height = 0.5in, anchor=center] (e2) at ([yshift=0.05in]trfig.center) {};
        \node[anchor=center] at (e2.180)  ()
                {\includegraphics[width=0.5in]{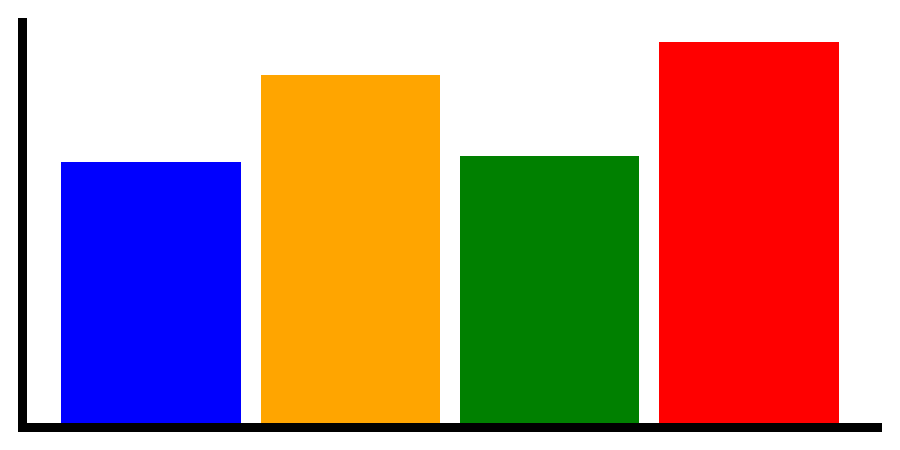}};
        \node[anchor=center] at ([yshift=-0.05in]e2.90)  ()
                {\includegraphics[width=0.4in]{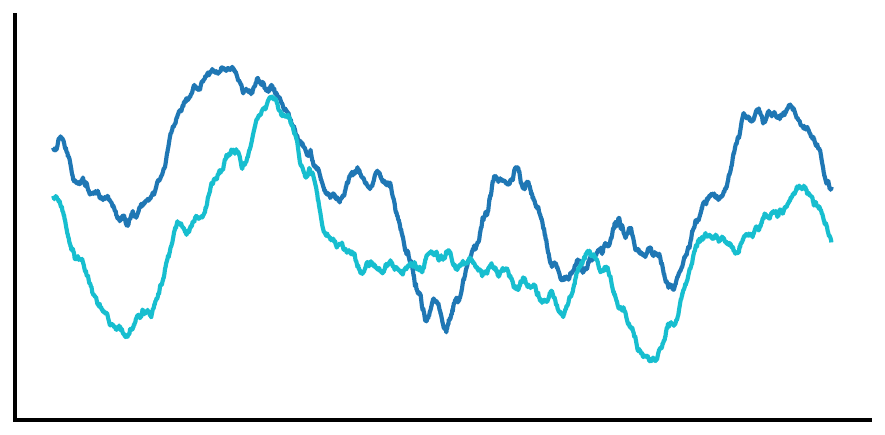}};
        \node[anchor=center] at (e2.0)  ()
                {\includegraphics[width=0.4in]{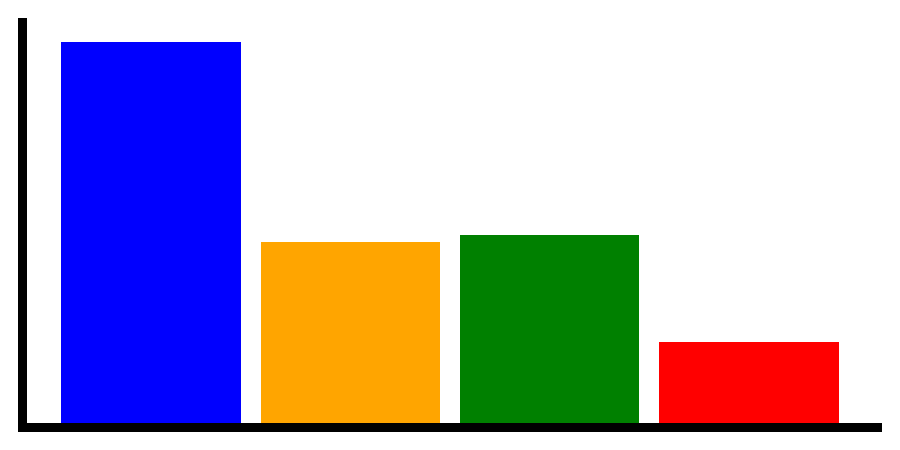}};
        \node[anchor=center] 
                at ([yshift=0.025in]e2.270)  (dcm1) {\includegraphics[height=0.35in]{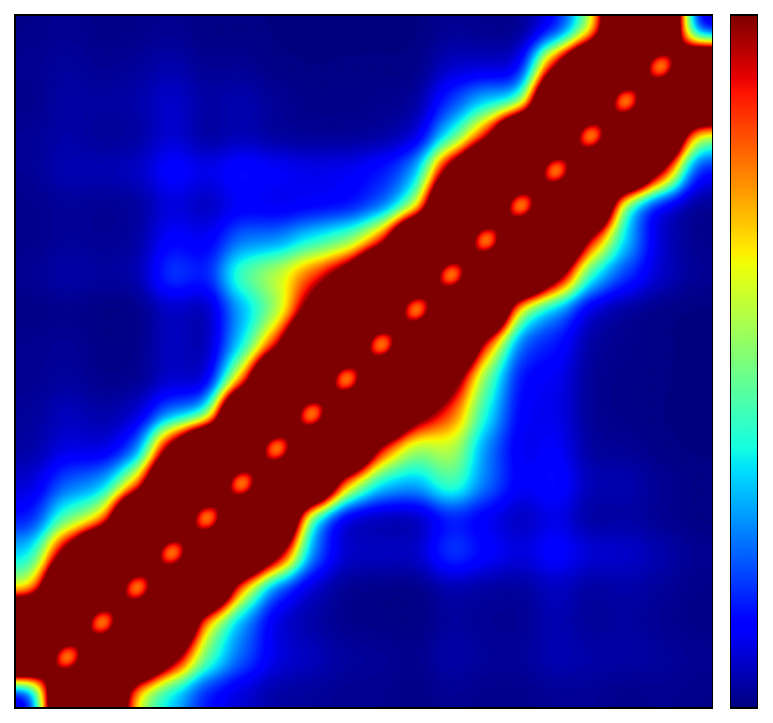}};
        
        \node [STEP, inner sep=0pt, minimum height=1.4in, minimum width=0.6in, text width=1in, align=center] at (13.5,1) (fanal) 
                {};./figures/Figure
        \node[draw=none, fill=none] at ([yshift=-0.15in]fanal.90) {\underline{Final  Analysis}} ;
        \node[rectangle, draw=none, fill=none, minimum width=0.5in, minimum height=0.5in] at ([yshift=-0.075in]fanal.center) (e3) {} ;
        \node[anchor=center] at (e3.135)  ()
                {\includegraphics[width=0.35in]{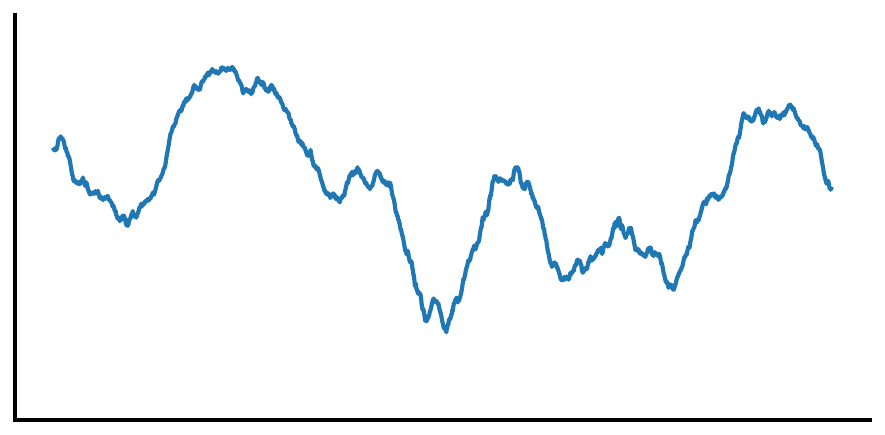}};
        \node[anchor=center] 
                at (e3.45)  (dcm0) {\includegraphics[width=0.4in]{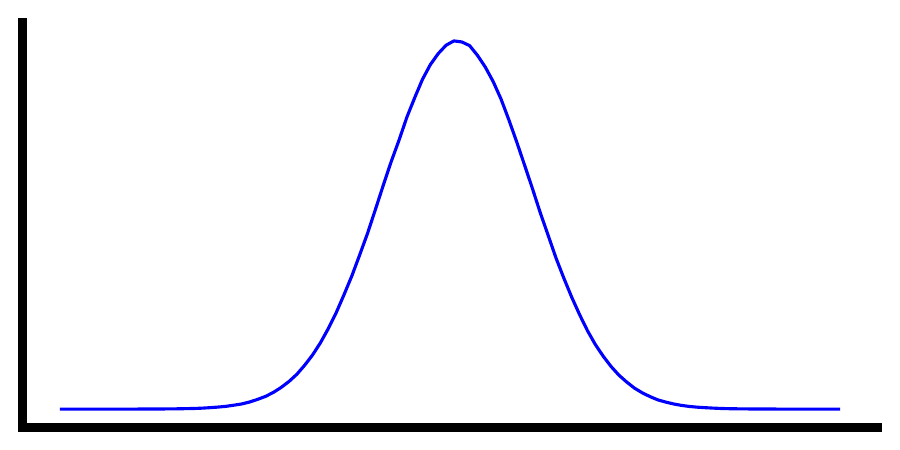}};
        \node[anchor=center] at (e3.315)  ()
                {\includegraphics[height=0.35in]{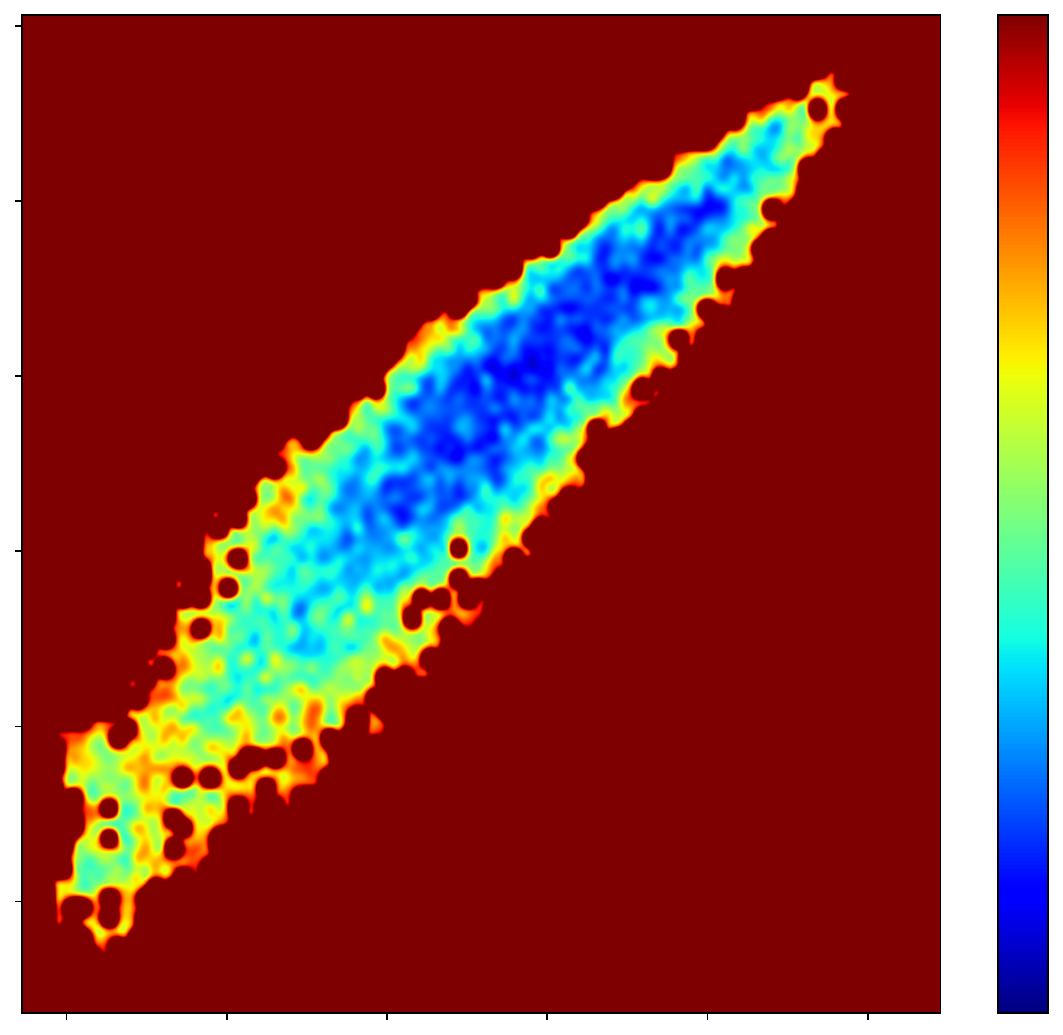}};
        \node[anchor=center] 
                at (e3.225)  () {\includegraphics[height=0.35in]{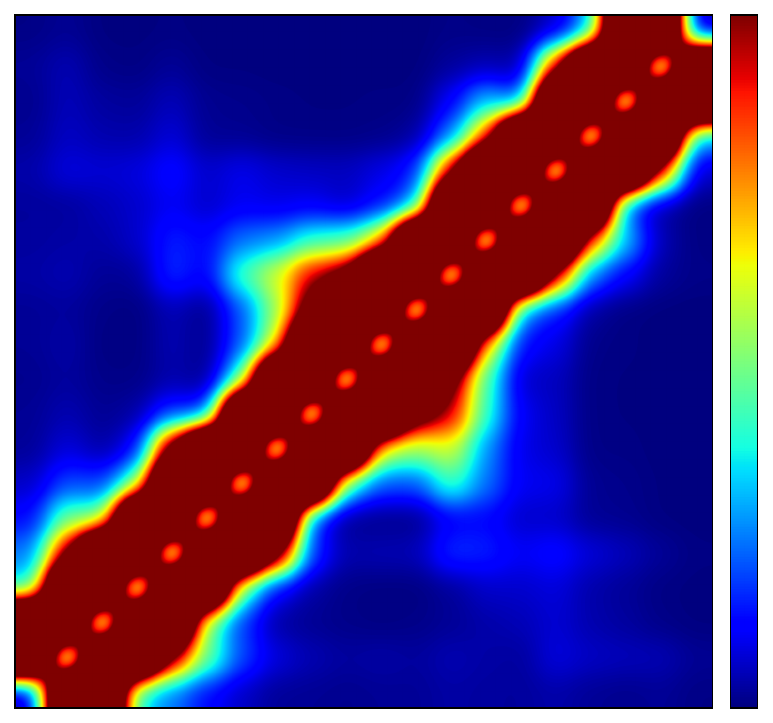}};

        \draw (temprep.east) edge[out=-7.5,in=180,-latex, line width=3] (trfig.west) node[draw=none,fill=none,font=\scriptsize, below, yshift=-0.06in, xshift=0.45in, text width=0.6in, align=center] {Temperature Replica Analysis};
        \draw (demuxrep.east) 
            edge[out=-30,in=180,-latex, line width=3] (dmfig.west) 
                node[draw=none,fill=none,font=\footnotesize, below, xshift=0.43in, yshift=-0.3in, text width=0.6in, align=center] 
                    {Demultiplexed Replica Analysis};
        \draw (dmfig.east) 
            edge[dashed, out=-70,in=70,latex-latex, line width=2] (trfig.east) 
                node[draw=none,fill=none,font=\footnotesize, below, xshift=0.275in, yshift=-0.45in, rotate=-90, text width=1.25in,align=center] 
                    {Assess Convergence};
        \draw 
            [loosely dashed,-latex, line width=2] (trfig.335) -|  
                node[draw=none,fill=none,font=\footnotesize, xshift=-0.65in, yshift=0.15in, text width=1.25in,align=center] 
                    {$\mathbf{300K}$ \textbf{Temperature Replica Trajectory} }(fanal.290) ;
      
    \end{tikzpicture}
    \caption{A workflow for preparing, running and analyzing REST2 simulations. a) Each box describes one of the four stages of preparing, running, and analyzing REST2 simulations. b) We depict the diffusion of replicas through a schematic solute temperature.  We identify the unscaled T = 300 K base temperature replica with a black box. Each colored line represents a demultiplexed replica diffusing through the solute temperature ladder. The center insets show cartoon representations of useful analyses to perform on temperature replicas and demultiplexed replicas to asses the convergence of REST2 simulations. After convergence analyses suggest statistically meaningful sampling of IDP conformational space, a final set of analyses is conduced on the unscaled T = 300 K base temperature replica. }\label{fig:workflow}
\end{figure}

An outline of the procedure to set--up, perform and analyze REST2 simulations of IDPs is displayed in Figure~\ref{fig:workflow}a. The four stages of running REST2 simulations are as follows. First, appropriate starting structures are chosen for each replica and simulation boxes are prepared by solvating each structure and adding the desired concentration of ions.
Second, equilibration simulations of each replica are performed to ensure that each simulation box has an appropriate pressure and volume. Poorly equilibrated replicas may result in low acceptance ratios between replicas and inefficient sampling. Third, one generates scaled topology files according to a desired solute temperature ladder and runs initial REST2 simulation steps to ensure that replicas are exchanging with desirable acceptance ratios before running a longer production simulation. Forth, a production REST2 simulation is performed. When a production REST2 simulation is running, it is important to intermittently analyze the conformational ensembles of each temperature rung and each demultiplexed replica, to ensure that no replicas have become "stuck" in the unscaled base temperature replica, as this can negate the efficiency of running REST2 (Figure~\ref{fig:workflow}b). These analyses allow one to check the conformational sampling in REST2 simulations, and assess the statistical convergence of structural properties of interest to determine if the simulation has been run for a sufficient length. In what follows, we discuss the details of each stage.

\subsection*{\textit{Producing Initial IDP Conformations for REST2 Simulations}}

Selecting appropriate initial structures is important for performing conventional MD simulations and REST2 simulations of IDPs. If nonphysical structures (such as structures with cis peptide bonds with dihedral angle $\omega=0^\circ$) or structures with very small Boltzmann weights (such as structures obtained from homology models resembling folded proteins) are selected as starting structures for IDP simulations, a substantial amount of simulation time may be wasted sampling largely irrelevant regions of conformational space. If IDP simulations are initiated from structures resembling folded proteins, considerable simulation time may be required to unfold these conformations before relevant regions of conformational space can be sampled. 

One possible option is to begin IDP simulations from extended linear chains, with backbone dihedral angles set to $\phi=-180^\circ$ and $\psi=+180^\circ$. Extended IDP conformations can be constructed using software such as pmx~\citep{Gapsys2015}, pymol~\citep{PyMOL} or Avogadro~\citep{Hanwell2012}. This may be a reasonable option for shorter IDPs and IDP fragments (10 to 40 residues), but for longer IDP chains, it may not be possible to insert a linear chain into a reasonably sized simulation box without obtaining contacts or clashes between a molecule and its periodic image. In these instances, short vacuum simulations of linear IDP chains can be performed to produce more collapsed conformations to use as starting structures in explicit solvent simulations. We note that care should be taken when preparing starting structures of IDPs with proline residues. Linear conformations are not well tolerated by prolines, and a polyproline II (ppII) helix (with $\Phi$=$-75^\circ$ and $\Psi$=$145^\circ$) may be a more reasonable set of initial dihedral angles for proline residues. We also caution that in most instances, unless experimental evidence indicates substantial populations of cis--proline conformations, one should set the peptide bond $\omega$ angle equal to the trans conformation ($\omega$=$180^\circ$). 

Many IDPs have substantial populations of helical secondary structure elements in solution that have been experimentally identified by NMR spectroscopy. If one has prior knowledge of the location of residual helical elements in an IDP, initializing simulations with starting structures with varied helical content in these regions may facilitate faster convergence to equilibrium averages. Adding helical elements into starting structures can also produce conformations with more realistic dimensions than linear chains. This 
 can be accomplished by setting the dihedral of residues with experimental helical propensities to the dihedral angles of canonical $\alpha$--helices ($\phi=-57^\circ$ and $\psi=-47^\circ$). Additional strategies for generation starting structures are discussed in Note~\ref{notes:secondary_structure}. 

We note that one can choose to start all replicas in REST2 from the same starting structure or to choose different starting structures for each replica. Using the same starting structure in each replica may increase the sampling time required to obtain robust statistics of structural properties of interest, as each replica may spend an appreciable portion of a REST2 simulation sampling the same region of conformational space near the starting structure, and it may difficult to detect a lack of convergence if no replicas discover new regions of conformational space in shorter REST2 simulations. REST2 simulations of IDPs that sample both helical and non-helical states may converge to equilibrium conformational distributions more quickly if a variety of starting structures that contain different residues in helical conformations and non-helical conformations are used as starting structures in different replicas. In the ergodic limit of infinite sampling, after an initial equilibration period of REST2 simulations are discarded, there should be no starting-structure dependence on computed equilibrium properties. In practice, however, ensemble properties of REST2 simulations run for 1--10 $\mu$s per replica will usually retain some starting--structure dependence, and it is therefore important to quantify statistical errors of simulated structural properties of interest. 

After selecting starting conformations for REST2 simulations, one needs to select an appropriate simulation box size, and the corresponding appropriate number of water models, for generating solvent boxes for all replicas. We note that each replica must contain the same number of water molecules and ions. If one utilizes linear IDP conformations as starting structures for simulations, traditional methods for estimating suitable box--sizes using solvent buffer regions may not produced useful estimates. A useful rule of thumb for identifying a suitable box--size for an IDP simulation is to use water--box with a box--length that is roughly 4 times the simulated radius of gyration ($R_{g}$) of an IDP ~\citep{shabane2019general}. IDP ensemble dimensions can be predicted using IDP ensemble property prediction methods~\citep{Lotthammer2024albatross} or using estimates based on the flory random coil model ~\citep{alston2023analytical}. Ultimately, as REST2 simulations will sample both extended and compact conformations of IDPs, one should check that a sufficiently large box--size is choosen to avoid frequent contacts with an IDP and its periodic image. Further disucssion is provided in Note~\ref{notes:box_size}. 

One must also decide how many replicas to use in a REST2 simulation. This is often a practical decision based on the computational resources available for a simulation. In practice, one should use as many replicas that can be simulated in parallel without decreasing the aggregate simulation time that can be achieved on a daily basis. For example, if a compute node has 10 GPUs, and simulations run with 1 replica per GPU have a performance of 150 ns/day while simulations run with 2 replicas per GPU each a have performance of 100 ns/day each, it is more efficient to run a REST2 simulation with 20 replicas. In this example, running with 20 replicas would produce 100 ns/day per replica, for an aggregate simulation performance of 2000 ns/per day across all replicas. A REST2 simulation run with 10 replicas on the same 10 GPUs, running at a speed of 150 ns/day per replica, would only produce an aggregate performance of 1500 ns/day.

In the accompanying tutorial, we simulate the last 20 residues of $\alpha$--synuclein (residues 121--140) with the following amino acid sequence: DMPVDPDNEAYEMPSEEGYQDYEPEA. We refer to this construct as $\alpha$--syn--C--term~\citep{Robustelli2022}. 
NMR data demonstrate that this region of $\alpha$--synuclein is highly disordered, with no appreciable populated secondary structure elements~\citep{Robustelli2018,Robustelli2022}. To begin the tutorial, first download the files from the accompanying GitHub repository:
\begin{lstlisting}[language=bash,xleftmargin=0mm,autogobble,breaklines=true]
 git clone https://github.com/paulrobustelli/IDP_REST_tutorial.git
\end{lstlisting}

In these files, we provide an $\alpha$--syn--C-term starting structure as well as example scripts to prepare an extended conformation of this sequence with the pmx python library~\citep{Gapsys2015}. We note that we use a starting structure without N--terminal or C--terminal capping groups. If one is simulating a fragment of an IDP that does not contain the natural termini of the protein, it is advised to consider appending terminal caps to prevent artifactual charge interactions from affecting the simulation results. N-terminal Acetyl--groups or C--terminal N--methyl amide--group caps can be applied to N--terminal or C--terminal residues of simulated constructs, respectively. Terminal caps can be added with the pymol~\citep{PyMOL} build function or with the pmx software~\citep{Gapsys2015}. 


\subsection*{\textit{Equilibrating Each Replica}}
After generating initial protein conformations for each replica, we next solvate each system with explicit water molecules and add ions, perform an energy minimization, and run simulations to equilibrate the pressure and box--size of each replica. If the same starting structure is going be used for each REST2 replica, a single equilibration can be performed. If different starting conformations are used for each replica, the simulation box of each starting conformation must be equilibrated seperately. The required equilibration steps are described below.

\begin{description}

    \item[Solvate and neutralize starting structures:] Starting conformations for simulations are placed in a simulation box with of specified box-length, and the simulation box is filled with solvent molecules and additional ions (e.g., $Na^{+}$, $Cl^{-}$). If the simulation is going to be compared with experimental data, it is advisable to perform the simulation with the same ion concentrations used in experiments. If the simulated IDP has a net charge, it is essential that ions are added such that the full system has a neutral charge. If different starting structures are used for each REST2 replica, each simulation box must have the same number of water molecules and same number of ions. 
    
    \item[Minimization:] Solvated replicas are subjected to an energy minimization step to remove unfavorable atomic contacts or strain introduced when adding solvent and ions or preparing starting structures.  

    \item[Heating:] After minimization, the system is gradually heated from a low temperature (often close to 0 K) to the target simulation temperature (e.g. 300 K).
    This step ensures the system is gently brought to the desired thermal conditions without introducing large energy fluctuations or instabilities.

    \item[Equilibration:] Replicas are equilibrated at a target temperature. If simulations will be run in the canonical (NVT) ensemble, it is important to ensure the pressure and box--size of each replica have equilibrated to stable values in the isobaric--isothermal (NPT) ensemble.

\end{description}

In the accompanying tutorial, initial conformations of $\alpha$--syn--C--term are placed in a 6.5 nm cubic box containg 8763 a99SB--\textit{disp} water molecules. $\alpha$--syn--C--term has a net charge of --8, and we add 8 $Na^{+}$ ions to achieve charge neutrality. In accordance with the a99SB--\textit{disp} guidelines~\citep{Robustelli2018} we utilize the CHARMM22 ions~\citep{MacKerell1998}. Details on determining the appropriate number of water molecules to use when solvating multiple different starting structures and GROMACS commands for solvation and charge neutralization are provided in Note \ref{notes:solvate}. 

The system is then minimized using up to 5000 steps of the steepest decent minimization algorithm. A successful minimization will relax the system to a maximum force of 100 $kJ mol^{-1} nm^{-1}$. To generate initial velocities we thermalize the simulation box by slowly heating the system to the desired temperature (300 K in our example) in the canonical ensemble (NVT) for about 1 ns. The ensemble is coupled to a temperature bath using V--rescale algorithm~\citep{Bussi2007}. The solvated system is equilibrated to 300 K. A brief 1 ns equilibration using the isobaric--isothermal (NPT) ensemble employing the Berendsen barostat~\citep{Berendsen1984} is used to bring the system pressure to 1 bar, relaxing the box volume and pressure. The Berendsen algorithm is computationally efficient, but does not produce accurate distributions of kinetic energy~\citep{Lemak1994,Morishita2000,Golo2002,Shirts2013}. After an initial relaxation, we then conduct a longer NPT equilibration with the more robust Parrinello--Rahman barostat ~\citep{parrinello1980crystal}. We equilibrate under NPT conditions for 40 ns with the V--rescale thermostat and Parrinello--Rahman barostat to achieve volume and pressure convergence at 300 K and 1 bar. It is important to select appropriate time constants for pressure and temperature coupling during this equilibration. We recommend 1 ps and 5 ps for temperature and pressure coupling times, respectively. We emphasize that as REST2 simulations will be performed in the canonical (NVT) ensemble, it is important that each replica is equilibrated to the desired system pressure, as swap attempts between replicas with large pressure differences can become extremely inefficient, causing sampling bottlenecks. NPT equilibration may need to be extended for some replicas if the desired pressure is not obtained. GROMACS simulation parameter (MDP) files for minimization and equilibration are provided in the accompanied online tutorial and GROMACS commands are described in Note \ref{notes:eq}. After checking for pressure and box-sie convergence, we use the final conformation of each NPT equilibration as the starting configurations for REST2 simulations.


\subsection*{\textit{Creating REST2 Simulation Input Files}}
After equilibrating simulation boxes for each replica, we generate topology files that contain the appropriate Hamiltonian scaling for each desired solute temperature in our REST2 temperature solute ladder. Scaled topology files are generated using PLUMED. We first generate a single unified processed topology file (processed.top) that contains all topology parameters for a replica in a single file. The GROMACS command to generate a unified processed topology file is contained in Note~\ref{notes:processed_topol}. If all replicas have the same number of protein, water and ion atoms, each replica will have the same unified processed topology file before Hamiltonian scaling.

This processed topology file must then be edited to indicate all atoms that should be considered as solute atoms in REST2 simulations. We edit processed.top to identify all atom types that will included in the solute region, and save this file as REST.top. Details of producing the REST.top file are discussed in Note~\ref{notes:processed_topol}. After generating the REST.top file that flags all solute atoms for Hamiltonian scaling, the PLUMED partial tempering script is used to read the REST.top topology file and output a scaled topology file that reflects the appropriate solute scaling factor for the desired solute temperature of each replica. 
As we will carry out each independent replica simulation in its own directory, we write out the appropriate scaled topology for each replica into the corresponding replica directory. The scaling factor supplied to the script is $\lambda_n=\frac{T_0}{T_n}$, where that $T_n$ is the solute temperature of the n$^{th}$ replica and $T_0$ is the base temperature replica ($T_0$ = 300 K here).  We first create directories for each of the n REST replicas (numbered \{0 1 2 3 ... n\}). In the following command, we deposit the scaled topology file (scaled.top) in the appropriate directory for the n$^{th}$ replica, \textbf{n}:

\begin{lstlisting}[language=sh, basicstyle=\ttfamily\small, breaklines=true,escapeinside={(*}{*)}]
    plumed partial_tempering   (*$\lambda_{n}$*) < REST.top > n/scaled.top
\end{lstlisting}
 
To perform parallel REST2 simulations, we must generate a GROMACS TPR file (a portable binary run input file) for each replica, and add a plumed.dat file into the simulation directory of each replica. First we copy each equilibrated structure file from the each NPT equilibration run and a copy of our production mdp file (prod.mdp), which contains the simulation parameters for production NVT runs, into the appropriate REST2 replica folder. Each replica folder now contains a scaled topology file for the selected temperature, the mdp file prod.mdp and an equilibrated structure file. In each replica simulation directory, we run the following command to produce a TPR for each replica and the plumed.dat file:
\begin{lstlisting}[language=sh, basicstyle=\ttfamily\small, breaklines=true]
    gmx grompp -f prod.mdp -c npt1.gro -p scaled.top -o scaled.tpr
    touch plumed.dat
\end{lstlisting}
Now each directory contains the two necessary files for running a REST2 simulation with the gromacs MD engine mdrun (scaled.tpr and plumed.dat). Running the REST2 the simulation is discussed in the following section. 

\subsection*{\textit{Running REST2 Simulations}}
When performing simulations on GPUs, GROMACS may automatically assign replicas across a number of available GPUs, or the user may specify computational resources to be used for simulatiosn of each replica. Because replicas will need to be time--synchronized to perform exchanges, it is logical to apply equivalent computational resources to each replica. For example, for a 20 replica simulation on 10 GPUs, one would assign two replicas to each GPU. 
For a 20 replica simulation on 5 GPUs, one would assign 4 replicas to each GPU.

After setting up each replica simulation directory, we run the simulation by adding the replica directories (ie. 1 2 3 4 5 ... n) into the following command:
\begin{lstlisting}[language=sh, basicstyle=\ttfamily\small, breaklines=true]
    mpirun -np 10 gmx mdrun -s scaled.tpr -multi <list of replica folders> -replex 800 -deffnm replica -plumed plumed.dat
\end{lstlisting}

This command specifies that replica swaps will be attempted every 800 simulation steps. We use a 2 $fs$ MD timestep, meaning that swaps are attempted every 1.6 $ps$. Shortly after a REST2 simulation is started, it is necessary to check the acceptance ratios of exchange attempts between replicas to ensure that there is sufficient energy overlap between solute temperature rungs in the selected temperature ladder to enable efficient exchanges. As a rule of thumb, it is desirable to use solute temperature ladder where swap attempts have a success rate of at least $\approx$20\%. If acceptance rates of swap attempts are 0 between some replicas, it is likely that either an error has been made when generating scaled topology files, or replicas must be spaced more closely together in the solute temperature ladder. As REST2 simulations progress, one should continue to periodically examine average acceptance ratios and perform additional analyses of the diffusion of each replica across the solute temperature replica to ensure that no replicas have become permanently stuck in a given temperature rung. 

\subsection*{\textit{Simulation Analysis}}\label{sec:Meth-Anal}

When performing REST2 simulations, it is important to intermittently perform analyses to assess the conformational sampling obtained in each solute temperature replica and each demultiplexed replica (Figure 1b). A temperature replica will consist of all frames sampled in given replica directory. If replica 0 reflects an unscaled base temperature of 300 K, and is run in replica folder 0, then all the conformations in the trajectory file in this folder correspond to the 300 K temperature replica. Alternatively, demultiplexed replicas track the continuous evolution of a set of coordinates as they diffuse through the solute temperature ladder. Analyzing demultiplexed replicas requires parsing history of swap attempts contained in simulation log files to identify the portions of each trajectory that correspond to each demultiplexed replica. Instructions for parsing trajectory log files to construct demultiplexed replicas are contain in Note~\ref{notes:demux}. Before analyzing conformational properties of temperature replicas or demultiplexed replicas, one must first perform periodic boundary condition (PBC) corrections in the same fashion as conventional MD simulations to enable trajectory visualization and ensure that the calculated structural properties do not suffer from PBC artifacts. GROMACS commands performing PBC corrections are contain in Note~\ref{notes:PBC}.

In the accompanying online tutorial we provide scripts to analyze each demultiplexed replica's diffusion through the solute temperature ladder. This includes calculating the probability each demultiplexed replica is found at the base temperature $T_0$, the mean effective solute temperature of each demultiplexed replica, and the average round--trip--time of each demultiplexed replica (the average number of simulation steps required for a replica to travel from the base temperature rung to the highest temperature rung and return to the base temperature rung). These analyses are shown in Figure 2. These results provide an example of excellent temperature diffusion of demultiplexed replicas. If any replicas became stuck in the base replica, this would be reflected by some replicas having very high probabilities of being found at $T_0$, a lower mean effective temperature, and a longer average round trip time. If one replica does become stuck at $T_0$, this can strongly influence the statistics of structural properties observed in this replica, and one should be cautious in interpreting these results.

\begin{figure}[!ht]
    \centering
    \includegraphics[width=0.6\linewidth]{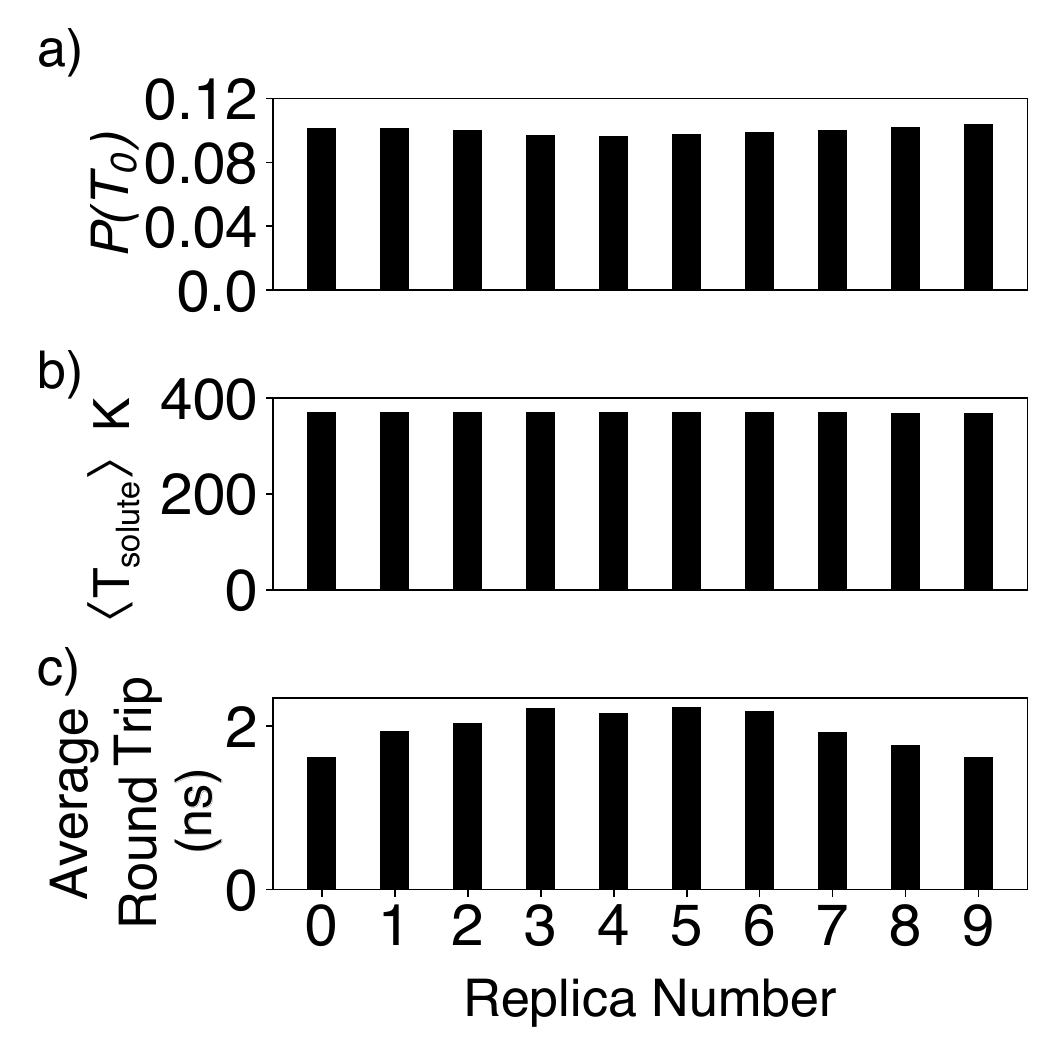}
    \caption{Demultiplexed replica temperature diffusion analysis. (a) Probability  that each demultiplexed replica is found at the base temperature 300 K. (b) Mean effective solute temperature sampled by each demultiplexed replica. (c) Average round trip time for each demultiplexed replica to travel from the lowest temperature rung to the highest temperature rung and return to the lowest temperature rung.}
    \label{fig:xvg_temp}
        
\end{figure}

    \begin{figure}[!ht]
    \begin{center}
        \begin{minipage}{4.5in}
            
            \includegraphics[width=\linewidth]{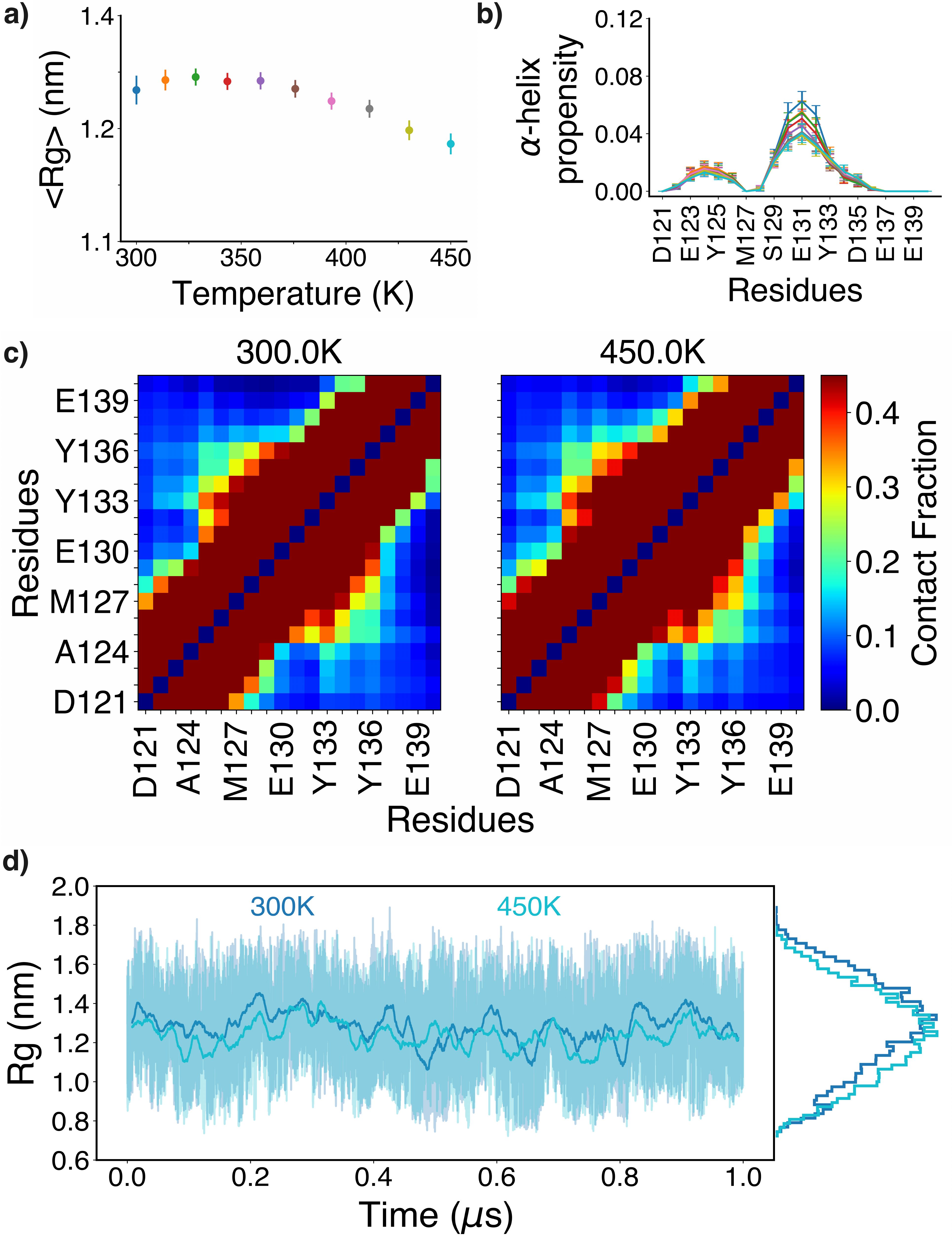}
            \caption{Analysis of the REST2 solute temperature replicas. We compare the 
            mean $R_{g}$ (a) and $\alpha$--helical content (b) of each temperature replica.              
            (c) Intramolecular contacts between amino acids of $\alpha$--syn--C--term for the base replica (T = 300 K) and the highest solute temperature (T = 450K) replica.
            (d) Time evolution of $R_g$ for two solute temperature replicas.  A window average is plotted in the foreground using a window length of 1.6 ns. Normalized histograms of the sampled $R_g$ values of each replica are displayed on the right of the plot.}
            \label{fig:rg_temp}
            
        \end{minipage}
        
    \end{center}
\end{figure}

Standard analysis of IDP simulations include calculating average secondary structure propensities, calculating the $R_g$ of each conformation and calculating the populations of intramolecular contacts. Scripts for calculating these properties are provided in the accompanying online tutorial. We compare the average structural properties of temperature replicas and demultiplexed replicas in Figure 3 and Figure 4, respectively. For each replica, we use a Flyvberg blocking analysis with the pyblock package to calculate statistical errors for each property of interest~\cite{Flyvbjerg1989,pyblock}. We note that for these structural properties, we expect to observe a temperature dependence across solute temperature replicas. In particular, with REST2 one frequently observes that at higher temperatures IDPs become more compact, form more intramolecular contacts, and contain less secondary structure (Figure 3). A smooth temperature dependence of each of these properties across temperature replicas as an indication of statistically meaningful conformational sampling. 

\begin{figure}[ht]
  \begin{center}
  
      \includegraphics[width=0.6\linewidth]{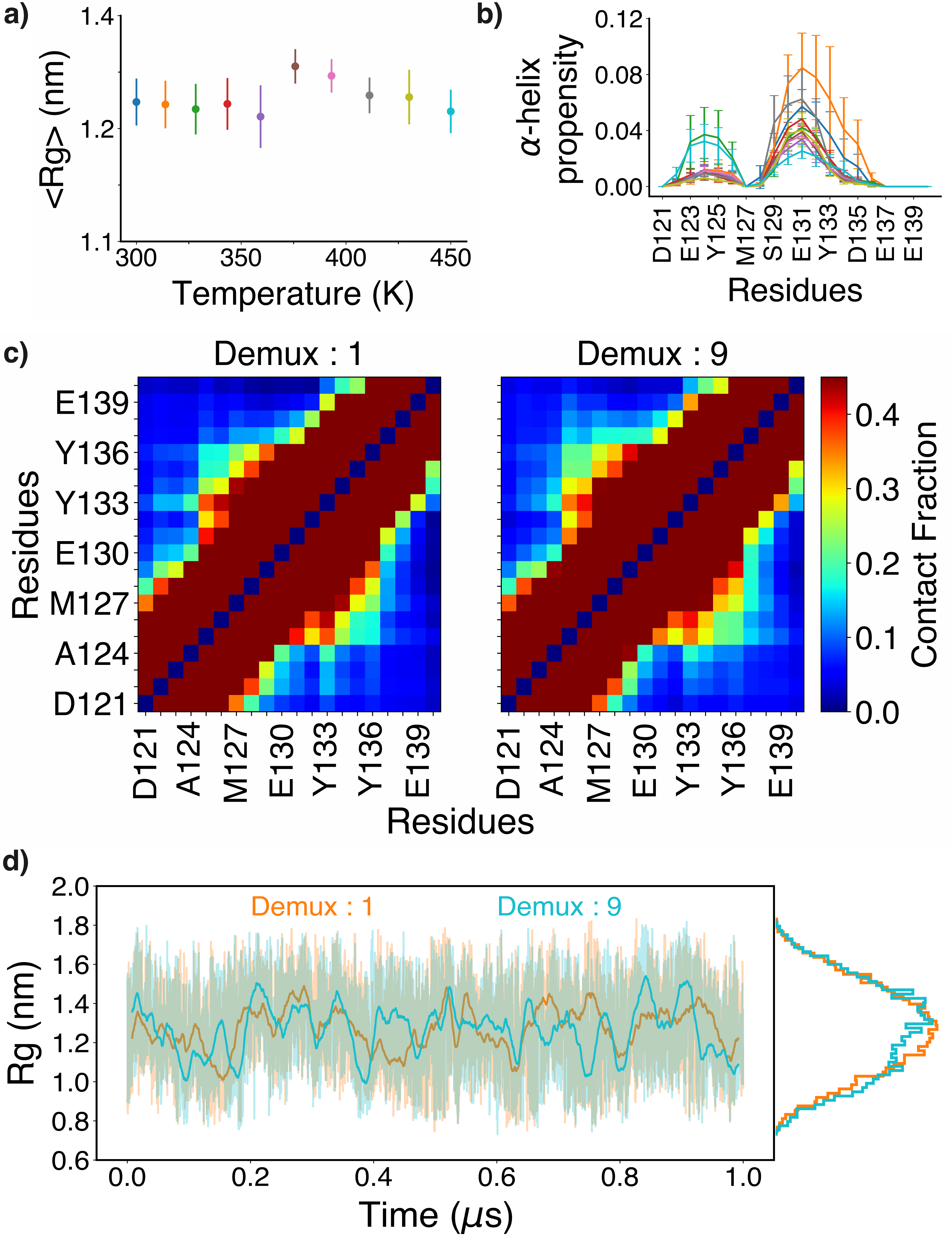}
      \caption{Analysis of the REST2 demultiplexed replicas. We compare the 
      mean $R_{g}$ (a) and $\alpha$--helical content (b) of each demultiplexed replica.      
      (c) Intramolecular contacts between amino acids of $\alpha$--syn--C--term for two demultiplexed replicas.
      (d) Time evolution of $R_g$ for two demultiplexed replicas. A window average is plotted in the foreground using a window length of 1.6 ns. Normalized histograms of the sampled $R_g$ values of each replica are displayed on the right of the plot.}
      \label{fig:rg_temp_d}
      
    
  \end{center}
\end{figure}

In a successful REST2 simulation, each demultiplexed replica will freely sample similar regions of conformational space, and in ergodic limit of infinite sampling across all temperature rungs, will converge to the same conformational distribution. In practice, due to finite sampling, each demultiplexed replica will have slightly different conformational properties. Ideally the average values observed across demultiplexed replicas will be within statistical sampling errors calculated by blocking. In practice, in longer IDP systems that are challenging to sample (ie. IDPs with more than 40 residues), some demultiplexed replicas will occasionally become stuck in free energy minima or sample different regions of conformational space. If the majority of demultiplexed replicas are sampling similar regions of conformational space and have similar conformational properties, however, it is likely that the statistical averages observed in the unscaled base temperature are statistically meaningful. 

In an unsuccessful, or uncoverged, REST2 simulation, many demultiplexed replicas will remain in their original conformational free energy basin, and sample only conformations similar to their starting structure. If every demultiplexed replicas samples a different region of conformational space, with little overlap, the REST2 simulation is clearly poorly converged, and the properties of the base temperature replicas are likely not meaningful. We note that in unsuccessful or uncoverged REST2 simulations, it may still be possible to achieve high acceptance rates in exchange attempts, as replicas may freely diffuse through the temperature ladder without undergoing major conformational rearrangements. Obtaining high acceptance rates on swap attempts is not, in isolation, sufficient evidence of robust conformational sampling or convergence. It is therefore essential to compare the distributions of conformations sampled in each demultiplexed replica to assess sampling efficiency and simulation convergence. 

\subsection*{Conclusion}\label{sec:Conclusion}
In this chapter we have introduced the theory behind replica exchange with solute tempering (REST) enhanced sampling methods and provided a practical guide to setting-up, performing and analyzing REST2 simulations of intrinsically disordered proteins (IDPs) using GROMACS and the PLUMED enhanced sampling plugin.  We have provided a detailed description of how to prepare initial conformations for REST2 simulations, solvate and equilibrate each replica, and prepare scaled REST2 topologies and simulation input files.   We describe how to analyze temperature replicas and independent demultiplexed temperature replicas from REST2 simulations and assess convergence of REST2 simulations.  We provide simulation input files and instructions for an accompanying tutorial GitHub repository:  \url{https://github.com/paulrobustelli/IDP_REST_tutorial}

\section*{Notes}

\begin{enumerate}
\item We provide instructions for a basic installation of plumed and gromacs with openmpi preinstalled on a Linux system with the gcc compiler in a bash environment. 
This series of commands assumes you have the variable \$MPICXX set. 
Further gains in performance can be reached by utilizing multiple GPUs and the CUDA compiler, nvcc, and CUDA runtime. Detailed descriptions on compiling GROMACS and PLUMED2 can be found at \href{http://gromacs.org}{http://gromacs.org} and \href{http:/plumed.org}{http:/plumed.org}, respectively.\\
        \begin{lstlisting}[language=bash,xleftmargin=0mm,autogobble,breaklines=true]
            #!/bin/bash
            INSTALL_ROOT=$HOME/opt
            mkdir src
            cd src
            git clone https://github.com/plumed/plumed2.git
            git clone https://github.com/gromacs/gromacs.git
            cd gromacs
            GMX_version=2024.3
            git checkout -b v$GMX_version$
            cd ../plumed2
            ./configure --prefix=$HOME/opt --enable-modules=all CXX="$MPICXX" CXXFLAGS="-O3 -axSSE2,AVX" 
            make
            make check
            make install
            echo 'export PLUMED_KERNEL="/usr/share/lib/libplumedKernel.so"' >> ~/.bashrc
            echo 'export PATH=$HOME/opt/bin:$PATH' >> ~/.bashrc
            echo 'export PLUMED_ROOT=$HOME/opt'
            source ~/.bashrc
            cd ../gromacs
            plumed patch -e gromacs-${GMX_version}
            mkdir build
            cd build
            cmake .. -DGMX_THREAD_MPI=OFF -DGMX_MPI=ON -DGMX_BUILD_OWN_FFTW=ON -DREGRESSIONTEST_DOWNLOAD=ON -DCMAKE_INSTALL_PREFIX=$HOME/opt
            make 
            make check 
            make install
        \end{lstlisting}\label{notes:installation}
    \item\label{notes:pythonEnv} We provide instructions for setting up a python environment for running the analysis contained in the accompanying online tutorial. This requires at minimum a miniconda environment installed prior to installing the subsequent packages. One can either install the necessary packages into the with the following command:
            \begin{lstlisting}[language=bash,xleftmargin=0mm,autogobble,breaklines=true]
                conda install -y matplotlib scipy mdtraj numpy MDAnalysis
                pip install pyblock
            \end{lstlisting}
            Or, within a python environment with pip3 installed, the following the necessary packages can be installed using the following command: 
            
            \begin{lstlisting}[language=bash,xleftmargin=0mm,autogobble,breaklines=true]
                pip install matplotlib scipy numpy mdtraj MDAnalysis pyblock
            \end{lstlisting}

    \item \label{notes:secondary_structure} 
    If experimental evidence indicates substantial populations of helical conformations in certain residues an IDP, or secondary structure propensity prediction algorithms (such as AGADIR~\citep{lacroix1998elucidating}) predict the presence of stable helices, one option is to generate an linear conformation and set the dihedral angles for residues with experimentally identified or computationally predicted helical conformations to canonical helical conformations ($\phi=-57^\circ$ and $\psi=-47^\circ$). One can then to perform a high temperature vacuum MD simulations to collapse the conformations into more realistic dimensions and gradually unfold helical elements, and choose various starting structures from these vacuum simulations spanning a range of helical content to use a starting structures in different REST2 replicas. 
    
    Alternatively, one can use a structure prediction of an IDP conformation from AlphaFold~\citep{Jumper2021} to generate largely extended, stereochemically validate starting structures of IDPs directly from sequence. AlphaFold predictions sometimes contain predicted regions of helical content, which can similarly be unfolded in short vacuum MD simulations to generate a range of starting structures for REST2 simulations. Alternatively, if there are experimental structures of an IDP bound to a binding--partner in the Protein Data Bank~\citep{Berman2000}, the coordinates of the IDP can be used as a starting conformations to generate initial conformations for REST2 simulations in an analogous fashion.  
    
    In all cases, before beginning IDP simulations, it is advisable to check the peptide--bond $\omega$ angle of all residues, to ensure that no cis--trans isomerizations have occurred in starting structure preparation. 
    
    \item \label{notes:box_size} When simulating a disordered protein, the protein will undergo expansion and compaction. A sufficiently large simulation box size should be selected to avoid nonphysical interactions self interactions between an IDP and its periodic image. After choosing a simulation box size, as a simulation progresses, it is advisable to check simulations for presence of periodic boundary contacts. A script (\textit{test\_pbc\_distances.ipynb}) is provided in the accompanying GitHub repository to calculate the minimum distance between periodic images of a protein. This script will flag all frames where the minimum distance between a protein and its periodic image is less than 2.0 nm. This analysis allows users to determine if the selected box size is appropriate for an IDP simulation. If the distance between protein atoms with its periodic image are substantially greater than 2.0 nm in all frames the resulting simulations box size can likely be reduced to enhance simulation speed. In contrast, if the simulation frequently contains periodic boundary contacts less than 2.0 nm the box size should be increased until only a small fraction of frames (ie. < 1\%) have contacts with their periodic image.  If a simulation has a substantial number of frames with PBC contacts, it is common to omit those frames from simulation analyses.

    \item \label{notes:solvate} The following commands generate a box of explicit solvent for a simulation using the --maxsol flag of the GROMACS gmx solvate function to ensure that each replica is solvated with the same number of water molecules:
            \begin{lstlisting}[language=bash, basicstyle=\ttfamily\small, breaklines=true]
                gmx pdb2gmx -f prot_only.pdb -o protein_processed.gro -ff a99sb_disp # make sure to have a99sb_disp.ff in the same folder
                gmx editconf -f protein_processed.pdb -o box.gro -box 6.5 6.5 6.5 # Box of length 6.5nm
                gmx solvate -cp box.gro -cs <a99sb_disp directory>/tip4pd.gro -p topol.top -maxsol 8763 -o solvate.gro
                gmx grompp -f minimz.mdp -c solvate.gro -p topol.top -o ions.tpr
                gmx genion -s ions.tpr -p topol.top -neutral -o solv_ions.gro -pname NA -nname CL # Select SOL
            \end{lstlisting}

    Here, we set the --maxsol parameter to 8763 water molecules, as this was found to fit into the chosen box of 6.5$nm$ for all starting structures. When performing REST2 simulations using different starting structures for each replica, each replica must contain the same number of water molecules and ions. To determine the most appropriate number of water molecules to use in all replicas, one can run test solvation calculations with a desired box size for each replica, identify the replica with fewest number of water molecules, and then build water boxes specifying this number of water molecules in each replica with gmx solvate --maxsol flag as demonstarted above. We note that if the selected box size does not fit the specified number of water molecules, a smaller number of water molecules may be added, and the process may need to be repeated with a larger box size.  It is therefore important to confirm that each replica contains the same number of water molecules after building each solvation box.

    \item \label{notes:eq} The following commands can be used for energy minimization and the different equilibration steps described in the main text for each replica:
                \begin{lstlisting}[language=bash, basicstyle=\ttfamily\small, breaklines=true]
                gmx grompp -f minimz.mdp -c solv_ions.gro -p topol.top -o asyn_minimz.tpr
                gmx mdrun -v -deffnm asyn_minimz
                gmx grompp -f NVT.mdp -c asyn_minimz.gro -p topol.top -o NVT.tpr
                gmx mdrun -v -deffnm NVT
                gmx grompp -f NPT0.mdp -c NVT.gro -p topol.top -o NPT0.tpr
                gmx mdrun -v -deffnm NPT0
                gmx grompp -f NPT1.mdp -c NPT0.gro -p topol.top -o NPT1.tpr
                gmx mdrun -v -deffnm NPT1
            \end{lstlisting}

     \item \label{notes:processed_topol} The following command will produce the united processed force field topology file (processed.top) required as input for selectively scaling the Hamiltonian for each REST2 replica with PLUMED:
                \begin{lstlisting}[language=bash, basicstyle=\ttfamily\small, breaklines=true]
                gmx grompp -f minimz.mdp -c solv_ions.gro -p topol.top -o temp.tpr -pp processed.top
                \end{lstlisting}
                
    Here we use the minimiz.mdp MDP file as input to generate the processed.top file, because it has no positional restraints. It is important to ensure that no positional restraints are included in the final processed.top that will be modified by PLUMED. 
    If an mdp file with positional restraints is used (for example NVT.mpd, which contains the line \textit{define = --DPOSRES}) a processed.top file will be written with positional restraints, and cannot be used to generate REST2 topology files. 

    \item \label{notes:scaling_atoms} The PLUMED \textit{partial\_tempering} script requires that an underscore "\_" be appended to the atomtype names of all solute atoms selected for  REST2 Hamiltonian scaling. For example, in the tutorial the first residue in the processed.top topology file is ASP. Under the section \textbf{[ atoms ]} the first line is:
            \begin{lstlisting}[language=bash, basicstyle=\ttfamily\small, breaklines=true]
                1         N3    121   NASP      N      1     0.0782      14.01
            \end{lstlisting}
    The second column is the atomtype name for this atom. The atomtype name needs to be appended with underscore as follows:
            \begin{lstlisting}[language=bash, basicstyle=\ttfamily\small, breaklines=true]
                1         N3_   121   NASP      N      1     0.0782      14.01
            \end{lstlisting}
            
    To facilitate easy appending of the underscore to the solute the following awk command will append underscores to the second column and output a REST.top file for a desired range of file numbers.        
            \begin{lstlisting}[language=awk, basicstyle=\ttfamily\small, breaklines=true]
                awk -v start_line=start -v end_line=end '{n++; if(n>start_line) if(n<end_line) if(NF>7) if($1+0==$1){$2=$2"_"}; print;}' processed.top > REST.top
            \end{lstlisting}
            
    The variables "start" and "end" specify the first and last line numbers of the processed.top file that contain atoms that will be selected as solute regions. These variables need to be substituted with the appropriate line numbers of the processed.top file. In the provided processed.top file of the accompanying online tutorial the protein solute atoms are contained between lines 964 and 1281. We therefore run the following awk command:

            \begin{lstlisting}[language=awk, basicstyle=\ttfamily\small, breaklines=true]
                awk -v start_line=964 -v end_line=1281 '{n++; if(n>start_line) if(n<end_line) if(NF>7) if($1+0==$1){$2=$2"_"}; print;}' processed.top > REST.top
            \end{lstlisting}

    \item \label{notes:demux} To obtain demultiplexed trajectories two xvg index files must be generated from the simulation log file. First create and enter a new sub--directory named demux. Now supply the log file of replica 0 to the perl script \textit{demux.fix.pl}:  

            \begin{lstlisting}[language=sh, basicstyle=\ttfamily\small, breaklines=true]
                perl demux.fix.pl ../0/production.log  
            \end{lstlisting}
    At the prompt enter the timestep chosen for the simulation, 0.002 ps for this tutorial. This produces two output files: \textit{replica\_index.xvg} and \textit{replica\_temp.xvg}. 
    
    \textit{replica\_temp.xvg} contains $(N_{replicas}+1)$ columns. The first column is the simulation time of exchange attempts, and the subsequent $N_{replicas}$ columns (one column for each demultiplexed replica) contains the index of the solute temperature rung that each demultiplixed replica populates at each simulation time. The accompanying online tutorial contains scripts to analyze \textit{replica\_temp.xvg} to compute the probability that each demultiplexed replica samples the base temperature replica, the average effective solute temperature of each demultiplexed replica, and the average simulation time required to complete a round trip from the base temperature replica, to the highest temperature replica, and back to the base temperature replica. These analyses are shown in Figure 2. 
    
    \textit{replica\_index.xvg} tracks the diffusion of demultiplexed replicas from the perspective of each temperature rung. The first column is the simulation time of exchange attempts. The subsequent $N_{replicas}$ columns each represent a temperature rung, and the value at each simulation timestep identifies the index of the demultiplexed replica present at a temperature rung at a given simulation time. This file is used to construct demultiplexed trajectories. If the timesteps written out in replica\_index.xvg are more frequent than the frequency with which coordinates are written out into trajectory files, the timeseries in replica\_index.xvg must be strided to match the frequency with which strucutres are saved. In the accompanying online tutorial, structures are saved every 80.0ps, while replica\_index.xvg contains an entry every 1.6ps. The following command appropriately strides the \textit{replica\_index.xvg} file to generate demultiplexed replicas:
    \begin{lstlisting}[language=awk, basicstyle=\ttfamily\small, breaklines=true]
        awk '{if($1==0$){print} if(n==50){$1=$1-80.0; print; n=0} n++;}' replica_index.xvg > replica_index.n50.s0.-80.xvg
    \end{lstlisting}
    The strided \textit{replica\_index.xvg} file, \textit{replica\_index.n50.s0.-80.xvg}, is then supplied to \textit{trjcat} following the \textit{--demux} flag along with the full path to the trajectory \textit{xtc} files and output to an equal number of demultiplexed \textit{xtc} trajectory files. The following \textit{trjcat} command provides all input files and output files including local directory location:
    \begin{lstlisting}[language=bash, basicstyle=\ttfamily\small, breaklines=true]
        gmx trjcat -demux replica_index.n50.s0.-80.xvg -f ../{0..9}/prod.xtc -o {0..9}.xtc
    \end{lstlisting}
    
  \item \label{notes:PBC}
  
  Periodic boundary condition (PBC) corrections are required to visualize and analyze trajectories. The following GROMACS commands can be performed to apply PBC corrections:

  \begin{lstlisting}[language=sh, basicstyle=\ttfamily\small, breaklines=true]
  gmx trjconv -f path/to/xtc/prod.xtc -s /path/to/tpr/prod.tpr -pbc nojump -o /path/to/nojump_out_file/nojump.xtc
  gmx trjconv -f /path/to/nojump_out_file/nojump.xtc -s /path/to/tpr/prod.tpr -pbc whole -o /path/to/whole_out_file/whole.xtc
  gmx trjconv -f /path/to/whole_out_file/whole.xtc -s /path/to/tpr/prod.tpr -pbc mol -o /path/to/pbc_file/pbc.xtc
\end{lstlisting}
   
\end{enumerate}

\section*{Acknowledgments}
This work was supported by the National Institutes of Health under award R35GM142750 (P.R., J.K.K.) and the Research Corporation for Science Advancement under Cottrell Scholar Award \#CS-CSA-2024-080 (K.M.R.) and the Dartmouth Innovation Accelerator for Cancer from the Magnuson Center for Entrepreneurship (K.M.R.).  P.R., J.K.K and K.M.R conceived and wrote the chapter.  We acknowledge Dr. Emanuele Scalone and Dr. Aparajita Charkaborty for helpful discussions.

\bibliography{enhanced_sampling.bib}

\begin{thebibliography}{60}
\providecommand{\natexlab}[1]{#1}
\providecommand{\url}[1]{\texttt{#1}}
\expandafter\ifx\csname urlstyle\endcsname\relax
  \providecommand{\doi}[1]{doi: #1}\else
  \providecommand{\doi}{doi: \begingroup \urlstyle{rm}\Url}\fi

\bibitem[Holehouse and Kragelund(2024)]{holehouse2024molecular}
Alex~S Holehouse and Birthe~B Kragelund.
\newblock The molecular basis for cellular function of intrinsically disordered protein regions.
\newblock \emph{Nature Reviews Molecular Cell Biology}, 25\penalty0 (3):\penalty0 187--211, 2024.

\bibitem[Banerjee et~al.(2023)Banerjee, Holehouse, Kriwacki, Robustelli, Jiang, Sobolevsky, Hurley, and Mendell]{banerjee2023dissecting}
Priya~R Banerjee, Alex~S Holehouse, Richard Kriwacki, Paul Robustelli, Hao Jiang, Alexander~I Sobolevsky, Jennifer~M Hurley, and Joshua~T Mendell.
\newblock Dissecting the biophysics and biology of intrinsically disordered proteins.
\newblock \emph{Trends in biochemical sciences}, 2023.

\bibitem[Bonomi et~al.(2017)Bonomi, Heller, Camilloni, and Vendruscolo]{bonomi2017principles}
Massimiliano Bonomi, Gabriella~T Heller, Carlo Camilloni, and Michele Vendruscolo.
\newblock Principles of protein structural ensemble determination.
\newblock \emph{Current opinion in structural biology}, 42:\penalty0 106--116, 2017.

\bibitem[Robustelli et~al.(2018)Robustelli, Piana, and Shaw]{Robustelli2018}
Paul Robustelli, Stefano Piana, and David~E. Shaw.
\newblock Developing a molecular dynamics force field for both folded and disordered protein states.
\newblock \emph{Proceedings of the National Academy of Sciences of the United States of America}, 115:\penalty0 E4758--E4766, 5 2018.

\bibitem[Sisk and Robustelli(2024)]{sisk2024folding}
Thomas~R Sisk and Paul Robustelli.
\newblock Folding-upon-binding pathways of an intrinsically disordered protein from a deep markov state model.
\newblock \emph{Proceedings of the National Academy of Sciences}, 121\penalty0 (6):\penalty0 e2313360121, 2024.

\bibitem[Zhu et~al.(2022)Zhu, Salvatella, and Robustelli]{zhu2022small}
Jiaqi Zhu, Xavier Salvatella, and Paul Robustelli.
\newblock Small molecules targeting the disordered transactivation domain of the androgen receptor induce the formation of collapsed helical states.
\newblock \emph{Nature Communications}, 13\penalty0 (1):\penalty0 6390, 2022.

\bibitem[Piana et~al.(2015{\natexlab{a}})Piana, Donchev, Robustelli, and Shaw]{Piana2015}
Stefano Piana, Alexander~G. Donchev, Paul Robustelli, and David~E. Shaw.
\newblock Water dispersion interactions strongly influence simulated structural properties of disordered protein states.
\newblock \emph{Journal of Physical Chemistry B}, 119:\penalty0 5113--5123, 2015{\natexlab{a}}.
\newblock \doi{10.1021/jp508971m}.

\bibitem[Huang et~al.(2017)Huang, Rauscher, Nawrocki, Ran, Feig, De~Groot, Grubm{\"u}ller, and MacKerell~Jr]{huang2017charmm36m}
Jing Huang, Sarah Rauscher, Grzegorz Nawrocki, Ting Ran, Michael Feig, Bert~L De~Groot, Helmut Grubm{\"u}ller, and Alexander~D MacKerell~Jr.
\newblock Charmm36m: an improved force field for folded and intrinsically disordered proteins.
\newblock \emph{Nature methods}, 14\penalty0 (1):\penalty0 71--73, 2017.

\bibitem[Best et~al.(2014)Best, Zheng, and Mittal]{best2014balanced}
Robert~B Best, Wenwei Zheng, and Jeetain Mittal.
\newblock Balanced protein--water interactions improve properties of disordered proteins and non-specific protein association.
\newblock \emph{Journal of chemical theory and computation}, 10\penalty0 (11):\penalty0 5113--5124, 2014.

\bibitem[Piana et~al.(2020)Piana, Robustelli, Tan, Chen, and Shaw]{Piana2020}
Stefano Piana, Paul Robustelli, Dazhi Tan, Songela Chen, and David~E. Shaw.
\newblock Development of a force field for the simulation of single-chain proteins and protein-protein complexes.
\newblock \emph{Journal of Chemical Theory and Computation}, 16:\penalty0 2494--2507, 4 2020.

\bibitem[H{\'e}nin et~al.(2022)H{\'e}nin, Leli{\'e}vre, Shirts, Valsson, and Delemotte]{henin2022enhanced}
J{\'e}r{\^o}me H{\'e}nin, Tony Leli{\'e}vre, Michael~R Shirts, Omar Valsson, and Lucie Delemotte.
\newblock Enhanced sampling methods for molecular dynamics simulations.
\newblock \emph{arXiv preprint arXiv:2202.04164}, 2022.

\bibitem[Sugita and Okamoto(1999)]{Sugita1999}
Yuji Sugita and Yuko Okamoto.
\newblock Replica-exchange molecular dynamics method for protein folding.
\newblock \emph{Chemical Physics Letters}, 314:\penalty0 141--151, 11 1999.

\bibitem[Hansmann(1997)]{hansmann1997parallel}
Ulrich~HE Hansmann.
\newblock Parallel tempering algorithm for conformational studies of biological molecules.
\newblock \emph{Chemical Physics Letters}, 281\penalty0 (1-3):\penalty0 140--150, 1997.

\bibitem[Liu et~al.(2005)Liu, Kim, Friesner, and Berne]{Liu2005}
Pu~Liu, Byungchan Kim, Richard~A. Friesner, and B.~J. Berne.
\newblock Replica exchange with solute tempering: A method for sampling biological systems in explicit water.
\newblock \emph{Proceedings of the National Academy of Sciences of the United States of America}, 102:\penalty0 13749--13754, 9 2005.
\newblock ISSN 00278424.

\bibitem[Wang et~al.(2011)Wang, Friesner, and Berne]{Wang2011}
Lingle Wang, Richard~A. Friesner, and B.~J. Berne.
\newblock Replica exchange with solute scaling: A more efficient version of replica exchange with solute tempering (rest2).
\newblock \emph{The Journal of Physical Chemistry B}, 115:\penalty0 9431--9438, 8 2011.
\newblock ISSN 1520-6106.

\bibitem[Bussi(2014)]{bussi2014hamiltonian}
Giovanni Bussi.
\newblock Hamiltonian replica exchange in gromacs: a flexible implementation.
\newblock \emph{Molecular Physics}, 112\penalty0 (3-4):\penalty0 379--384, 2014.

\bibitem[Spoel et~al.(2005)Spoel, Lindahl, Hess, Groenhof, Mark, and Berendsen]{VanDerSpoel2005}
David Van~Der Spoel, Erik Lindahl, Berk Hess, Gerrit Groenhof, Alan~E Mark, and Herman J~C Berendsen.
\newblock Gromacs: fast, flexible, and free.
\newblock \emph{Journal of computational chemistry}, 26:\penalty0 1701--18, 12 2005.
\newblock ISSN 0192-8651.

\bibitem[Abraham et~al.(2015)Abraham, Murtola, Schulz, P{\'a}ll, Smith, Hess, and Lindahl]{abraham2015gromacs}
Mark~James Abraham, Teemu Murtola, Roland Schulz, Szil{\'a}rd P{\'a}ll, Jeremy~C Smith, Berk Hess, and Erik Lindahl.
\newblock Gromacs: High performance molecular simulations through multi-level parallelism from laptops to supercomputers.
\newblock \emph{SoftwareX}, 1:\penalty0 19--25, 2015.

\bibitem[Bonomi et~al.(2009)Bonomi, Branduardi, Bussi, Camilloni, Provasi, Raiteri, Donadio, Marinelli, Pietrucci, Broglia, and Parrinello]{Bonomi2009}
Massimiliano Bonomi, Davide Branduardi, Giovanni Bussi, Carlo Camilloni, Davide Provasi, Paolo Raiteri, Davide Donadio, Fabrizio Marinelli, Fabio Pietrucci, Ricardo~A. Broglia, and Michele Parrinello.
\newblock Plumed: A portable plugin for free-energy calculations with molecular dynamics.
\newblock \emph{Computer Physics Communications}, 180:\penalty0 1961--1972, 10 2009.
\newblock ISSN 0010-4655.
\newblock \doi{10.1016/J.CPC.2009.05.011}.

\bibitem[Tribello et~al.(2014)Tribello, Bonomi, Branduardi, Camilloni, and Bussi]{Tribello2014}
Gareth~A. Tribello, Massimiliano Bonomi, Davide Branduardi, Carlo Camilloni, and Giovanni Bussi.
\newblock Plumed 2: New feathers for an old bird.
\newblock \emph{Computer Physics Communications}, 185:\penalty0 604--613, 2 2014.
\newblock ISSN 0010-4655.
\newblock \doi{10.1016/J.CPC.2013.09.018}.

\bibitem[Kamiya and Sugita(2018)]{kamiya2018flexible}
Motoshi Kamiya and Yuji Sugita.
\newblock Flexible selection of the solute region in replica exchange with solute tempering: Application to protein-folding simulations.
\newblock \emph{The Journal of chemical physics}, 149\penalty0 (7), 2018.

\bibitem[Lao et~al.(2024)Lao, O'Connor, and Huang]{lao2024replica}
Yichong Lao, Michael O'Connor, and Xuhui Huang.
\newblock Replica exchange with solute tempering for protein conformational sampling.
\newblock \emph{Journal of Chemical Theory and Computation}, 2024.

\bibitem[Appadurai et~al.(2021)Appadurai, Nagesh, and Srivastava]{Appadurai2021}
Rajeswari Appadurai, Jayashree Nagesh, and Anand Srivastava.
\newblock High resolution ensemble description of metamorphic and intrinsically disordered proteins using an efficient hybrid parallel tempering scheme.
\newblock \emph{Nature Communications 2021 12:1}, 12:\penalty0 1--11, 2 2021.
\newblock \doi{10.1038/s41467-021-21105-7}.

\bibitem[Zhang et~al.(2023)Zhang, Liu, and Chen]{Zhang2023}
Yumeng Zhang, Xiaorong Liu, and Jianhan Chen.
\newblock Re-balancing replica exchange with solute tempering for sampling dynamic protein conformations.
\newblock \emph{Journal of Chemical Theory and Computation}, 19:\penalty0 1602--1614, 3 2023.

\bibitem[Robustelli et~al.(2022)Robustelli, Ibanez-De-Opakua, Campbell-Bezat, Giordanetto, Becker, Zweckstetter, Pan, and Shaw]{Robustelli2022}
Paul Robustelli, Alain Ibanez-De-Opakua, Cecily Campbell-Bezat, Fabrizio Giordanetto, Stefan Becker, Markus Zweckstetter, Albert~C. Pan, and David~E. Shaw.
\newblock Molecular basis of small-molecule binding to $\alpha$-synuclein.
\newblock \emph{Journal of the American Chemical Society}, 144:\penalty0 2501--2510, 2 2022.
\newblock \doi{10.1021/JACS.1C07591}.

\bibitem[Bussi et~al.(2007)Bussi, Donadio, Parrinello, and Phys]{Bussi2007}
Giovanni Bussi, Davide Donadio, Michele Parrinello, and J~Chem Phys.
\newblock Canonical sampling through velocity rescaling.
\newblock \emph{J. Chem. Phys}, 126:\penalty0 14101, 2007.
\newblock \doi{10.1063/1.2408420}.
\newblock URL \url{https://doi.org/10.1063/1.2408420}.

\bibitem[Huang et~al.(2007)Huang, Hagen, Kim, Friesner, Zhou, and Berne]{huang2007replica}
Xuhui Huang, Morten Hagen, Byungchan Kim, Richard~A Friesner, Ruhong Zhou, and Bruce~J Berne.
\newblock Replica exchange with solute tempering: efficiency in large scale systems.
\newblock \emph{The Journal of Physical Chemistry B}, 111\penalty0 (19):\penalty0 5405--5410, 2007.

\bibitem[Prakash et~al.(2011)Prakash, Barducci, and Parrinello]{prakash2011replica}
Meher~K Prakash, Alessandro Barducci, and Michele Parrinello.
\newblock Replica temperatures for uniform exchange and efficient roundtrip times in explicit solvent parallel tempering simulations.
\newblock \emph{Journal of chemical theory and computation}, 7\penalty0 (7):\penalty0 2025--2027, 2011.

\bibitem[P\'{a}ll et~al.(2020)P\'{a}ll, Zhmurov, Bauer, Abraham, Lundborg, Gray, Hess, and Lindahl]{pall2020heterogeneous}
Szil\'{a}rd P\'{a}ll, Artem Zhmurov, Paul Bauer, Mark Abraham, Magnus Lundborg, Alan Gray, Berk Hess, and Erik Lindahl.
\newblock Heterogeneous parallelization and acceleration of molecular dynamics simulations in gromacs.
\newblock \emph{The Journal of Chemical Physics}, 153\penalty0 (13), 2020.

\bibitem[Humphrey et~al.(1996)Humphrey, Dalke, and Schulten]{HUMP96}
William Humphrey, Andrew Dalke, and Klaus Schulten.
\newblock {VMD} -- {V}isual {M}olecular {D}ynamics.
\newblock \emph{Journal of Molecular Graphics}, 14:\penalty0 33--38, 1996.

\bibitem[Schr\"odinger(2015)]{PyMOL}
LLC Schr\"odinger.
\newblock The {PyMOL} molecular graphics system, version~1.8.
\newblock Available at: \url{https://pymol.org/}, November 2015.

\bibitem[Yuan et~al.(2017)Yuan, Chan, and Hu]{Yuan2017}
Shuguang Yuan, H.~C.Stephen Chan, and Zhenquan Hu.
\newblock Using pymol as a platform for computational drug design.
\newblock \emph{Wiley Interdisciplinary Reviews: Computational Molecular Science}, 7:\penalty0 e1298, 3 2017.
\newblock ISSN 1759-0884.
\newblock \doi{10.1002/WCMS.1298}.
\newblock URL \url{https://onlinelibrary.wiley.com/doi/full/10.1002/wcms.1298 https://onlinelibrary.wiley.com/doi/abs/10.1002/wcms.1298 https://wires.onlinelibrary.wiley.com/doi/10.1002/wcms.1298}.

\bibitem[Van~Rossum and Drake(2009)]{python3}
Guido Van~Rossum and Fred~L. Drake.
\newblock \emph{Python 3 Reference Manual}.
\newblock CreateSpace, Scotts Valley, CA, 2009.
\newblock ISBN 1441412697.

\bibitem[Team(2024)]{matplotlibdev2024}
The Matplotlib~Development Team.
\newblock Matplotlib: Visualization with python, August 2024.

\bibitem[Hunter(2007)]{Hunter2007}
J.~D. Hunter.
\newblock Matplotlib: A 2d graphics environment.
\newblock \emph{Computing in Science \& Engineering}, 9\penalty0 (3):\penalty0 90--95, 2007.
\newblock \doi{10.1109/MCSE.2007.55}.

\bibitem[Harris et~al.(2020)Harris, Millman, van~der Walt, Gommers, Virtanen, Cournapeau, Wieser, Taylor, Berg, Smith, Kern, Picus, Hoyer, van Kerkwijk, Brett, Haldane, del R{\'{i}}o, Wiebe, Peterson, G{\'{e}}rard-Marchant, Sheppard, Reddy, Weckesser, Abbasi, Gohlke, and Oliphant]{harris2020array}
Charles~R. Harris, K.~Jarrod Millman, St{\'{e}}fan~J. van~der Walt, Ralf Gommers, Pauli Virtanen, David Cournapeau, Eric Wieser, Julian Taylor, Sebastian Berg, Nathaniel~J. Smith, Robert Kern, Matti Picus, Stephan Hoyer, Marten~H. van Kerkwijk, Matthew Brett, Allan Haldane, Jaime~Fern{\'{a}}ndez del R{\'{i}}o, Mark Wiebe, Pearu Peterson, Pierre G{\'{e}}rard-Marchant, Kevin Sheppard, Tyler Reddy, Warren Weckesser, Hameer Abbasi, Christoph Gohlke, and Travis~E. Oliphant.
\newblock Array programming with {NumPy}.
\newblock \emph{Nature}, 585\penalty0 (7825):\penalty0 357--362, Sep 2020.
\newblock \doi{10.1038/s41586-020-2649-2}.

\bibitem[Virtanen et~al.(2020)Virtanen, Gommers, Oliphant, Haberland, Reddy, Cournapeau, Burovski, Peterson, Weckesser, Bright, {van der Walt}, Brett, Wilson, Millman, Mayorov, Nelson, Jones, Kern, Larson, Carey, Polat, Feng, Moore, {VanderPlas}, Laxalde, Perktold, Cimrman, Henriksen, Quintero, Harris, Archibald, Ribeiro, Pedregosa, {van Mulbregt}, and {SciPy 1.0 Contributors}]{2020SciPy}
Pauli Virtanen, Ralf Gommers, Travis~E. Oliphant, Matt Haberland, Tyler Reddy, David Cournapeau, Evgeni Burovski, Pearu Peterson, Warren Weckesser, Jonathan Bright, St{\'e}fan~J. {van der Walt}, Matthew Brett, Joshua Wilson, K.~Jarrod Millman, Nikolay Mayorov, Andrew R.~J. Nelson, Eric Jones, Robert Kern, Eric Larson, C~J Carey, {\.I}lhan Polat, Yu~Feng, Eric~W. Moore, Jake {VanderPlas}, Denis Laxalde, Josef Perktold, Robert Cimrman, Ian Henriksen, E.~A. Quintero, Charles~R. Harris, Anne~M. Archibald, Ant{\^o}nio~H. Ribeiro, Fabian Pedregosa, Paul {van Mulbregt}, and {SciPy 1.0 Contributors}.
\newblock {{SciPy} 1.0: Fundamental Algorithms for Scientific Computing in Python}.
\newblock \emph{Nature Methods}, 17:\penalty0 261--272, 2020.
\newblock \doi{10.1038/s41592-019-0686-2}.

\bibitem[McGibbon et~al.(2015)McGibbon, Beauchamp, Harrigan, Klein, Swails, Hernández, Schwantes, Wang, Lane, and Pande]{McGibbon2015}
Robert~T. McGibbon, Kyle~A. Beauchamp, Matthew~P. Harrigan, Christoph Klein, Jason~M. Swails, Carlos~X. Hernández, Christian~R. Schwantes, Lee~Ping Wang, Thomas~J. Lane, and Vijay~S. Pande.
\newblock Mdtraj: A modern open library for the analysis of molecular dynamics trajectories.
\newblock \emph{Biophysical Journal}, 109:\penalty0 1528--1532, 10 2015.
\newblock ISSN 0006-3495.
\newblock \doi{10.1016/J.BPJ.2015.08.015}.

\bibitem[Michaud-Agrawal et~al.(2011)Michaud-Agrawal, Denning, Woolf, and Beckstein]{Michaud-Agrawal2011}
Naveen Michaud-Agrawal, Elizabeth~J. Denning, Thomas~B. Woolf, and Oliver Beckstein.
\newblock Mdanalysis: A toolkit for the analysis of molecular dynamics simulations.
\newblock \emph{Journal of Computational Chemistry}, 32:\penalty0 2319--2327, 7 2011.
\newblock \doi{10.1002/jcc.21787}.

\bibitem[Gowers et~al.(2016)Gowers, Linke, Barnoud, Reddy, Melo, Seyler, Domański, Dotson, Buchoux, Kenney, and Beckstein]{Gowers2016}
Richard Gowers, Max Linke, Jonathan Barnoud, Tyler Reddy, Manuel Melo, Sean Seyler, Jan Domański, David Dotson, Sébastien Buchoux, Ian Kenney, and Oliver Beckstein.
\newblock Mdanalysis: A python package for the rapid analysis of molecular dynamics simulations.
\newblock In \emph{PROC. OF THE 15th PYTHON IN SCIENCE CONF}, pages 98--105, 2016.
\newblock \doi{10.25080/Majora-629e541a-00e}.

\bibitem[Flyvbjerg and Petersen(1989)]{Flyvbjerg1989}
H.~Flyvbjerg and H.~G. Petersen.
\newblock Error estimates on averages of correlated data.
\newblock \emph{The Journal of Chemical Physics}, 91:\penalty0 461--466, 7 1989.
\newblock ISSN 0021-9606.
\newblock \doi{10.1063/1.457480}.

\bibitem[Spencer(2020)]{pyblock}
James Spencer.
\newblock pyblock: A python module for performing a reblocking analysis on serially-correlated data, 2020.
\newblock URL \url{http://github.com/jsspencer/pyblock}.

\bibitem[Paul~Bauer and Lindahl(2023)]{Gromacs2022.5}
Berk~Hess Paul~Bauer and Erik Lindahl.
\newblock Gromacs 2022.5 source code (2022.5), 2023.

\bibitem[Huang and MacKerell(2018)]{Huang2018}
Jing Huang and Alexander~D MacKerell.
\newblock Force field development and simulations of intrinsically disordered proteins.
\newblock \emph{Current Opinion in Structural Biology}, 48:\penalty0 40--48, 2 2018.
\newblock \doi{10.1016/j.sbi.2017.10.008}.

\bibitem[Piana et~al.(2015{\natexlab{b}})Piana, Donchev, Robustelli, and Shaw]{piana2015water}
Stefano Piana, Alexander~G Donchev, Paul Robustelli, and David~E Shaw.
\newblock Water dispersion interactions strongly influence simulated structural properties of disordered protein states.
\newblock \emph{The journal of physical chemistry B}, 119\penalty0 (16):\penalty0 5113--5123, 2015{\natexlab{b}}.

\bibitem[Gapsys et~al.(2015)Gapsys, Michielssens, Seeliger, and de~Groot]{Gapsys2015}
Vytautas Gapsys, Servaas Michielssens, Daniel Seeliger, and Bert~L. de~Groot.
\newblock pmx: Automated protein structure and topology generation for alchemical perturbations.
\newblock \emph{Journal of Computational Chemistry}, 36:\penalty0 348--354, 2 2015.
\newblock \doi{10.1002/jcc.23804}.

\bibitem[Hanwell et~al.(2012)Hanwell, Curtis, Lonie, Vandermeersch, Zurek, and Hutchison]{Hanwell2012}
Marcus~D Hanwell, Donald~E Curtis, David~C Lonie, Tim Vandermeersch, Eva Zurek, and Geoffrey~R Hutchison.
\newblock Avogadro: an advanced semantic chemical editor, visualization, and analysis platform.
\newblock \emph{Journal of Cheminformatics}, 4:\penalty0 17, 12 2012.
\newblock \doi{10.1186/1758-2946-4-17}.

\bibitem[Shabane et~al.(2019)Shabane, Izadi, and Onufriev]{shabane2019general}
Parviz~Seifpanahi Shabane, Saeed Izadi, and Alexey~V Onufriev.
\newblock General purpose water model can improve atomistic simulations of intrinsically disordered proteins.
\newblock \emph{Journal of chemical theory and computation}, 15\penalty0 (4):\penalty0 2620--2634, 2019.

\bibitem[Lotthammer et~al.(2024)Lotthammer, Ginell, Griffith, Emenecker, and Holehouse]{Lotthammer2024albatross}
Jeffrey~M. Lotthammer, Garrett~M. Ginell, Daniel Griffith, Ryan~J. Emenecker, and Alex~S. Holehouse.
\newblock Direct prediction of intrinsically disordered protein conformational properties from sequence.
\newblock \emph{Nature Methods}, 21:\penalty0 465--476, 3 2024.
\newblock \doi{10.1038/s41592-023-02159-5}.

\bibitem[Alston et~al.(2023)Alston, Ginell, Soranno, and Holehouse]{alston2023analytical}
Jhullian~J Alston, Garrett~M Ginell, Andrea Soranno, and Alex~S Holehouse.
\newblock The analytical flory random coil is a simple-to-use reference model for unfolded and disordered proteins.
\newblock \emph{The Journal of Physical Chemistry B}, 127\penalty0 (21):\penalty0 4746--4760, 2023.

\bibitem[MacKerell et~al.(1998)MacKerell, Bashford, Bellott, Dunbrack, Evanseck, Field, Fischer, Gao, Guo, Ha, Joseph-McCarthy, Kuchnir, Kuczera, Lau, Mattos, Michnick, Ngo, Nguyen, Prodhom, Reiher, Roux, Schlenkrich, Smith, Stote, Straub, Watanabe, Wiórkiewicz-Kuczera, Yin, and Karplus]{MacKerell1998}
A.~D. MacKerell, D.~Bashford, M.~Bellott, R.~L. Dunbrack, J.~D. Evanseck, M.~J. Field, S.~Fischer, J.~Gao, H.~Guo, S.~Ha, D.~Joseph-McCarthy, L.~Kuchnir, K.~Kuczera, F.~T.~K. Lau, C.~Mattos, S.~Michnick, T.~Ngo, D.~T. Nguyen, B.~Prodhom, W.~E. Reiher, B.~Roux, M.~Schlenkrich, J.~C. Smith, R.~Stote, J.~Straub, M.~Watanabe, J.~Wiórkiewicz-Kuczera, D.~Yin, and M.~Karplus.
\newblock All-atom empirical potential for molecular modeling and dynamics studies of proteins.
\newblock \emph{The Journal of Physical Chemistry B}, 102:\penalty0 3586--3616, 4 1998.
\newblock \doi{10.1021/jp973084f}.

\bibitem[Berendsen et~al.(1984)Berendsen, Postma, van Gunsteren, DiNola, and Haak]{Berendsen1984}
H.~J.~C. Berendsen, J.~P.~M. Postma, W.~F. van Gunsteren, A.~DiNola, and J.~R. Haak.
\newblock Molecular dynamics with coupling to an external bath.
\newblock \emph{The Journal of Chemical Physics}, 81:\penalty0 3684--3690, 10 1984.
\newblock \doi{10.1063/1.448118}.

\bibitem[Lemak and Balabaev(1994)]{Lemak1994}
A.~S. Lemak and N.~K. Balabaev.
\newblock On the berendsen thermostat.
\newblock \emph{Molecular Simulation}, 13:\penalty0 177--187, 9 1994.
\newblock \doi{10.1080/08927029408021981}.

\bibitem[Morishita(2000)]{Morishita2000}
Tetsuya Morishita.
\newblock Fluctuation formulas in molecular-dynamics simulations with the weak coupling heat bath.
\newblock \emph{The Journal of Chemical Physics}, 113:\penalty0 2976--2982, 8 2000.
\newblock \doi{10.1063/1.1287333}.

\bibitem[Golo and Sha\u{i}tan(2002)]{Golo2002}
V~L Golo and K~V Sha\u{i}tan.
\newblock Dynamic attractor for the berendsen thermostat an the slow dynamics of biomacromolecules.
\newblock \emph{Biofizika}, 47:\penalty0 611--7, 2002.

\bibitem[Shirts(2013)]{Shirts2013}
Michael~R. Shirts.
\newblock Simple quantitative tests to validate sampling from thermodynamic ensembles.
\newblock \emph{Journal of Chemical Theory and Computation}, 9:\penalty0 909--926, 2 2013.
\newblock \doi{10.1021/ct300688p}.

\bibitem[Parrinello and Rahman(1980)]{parrinello1980crystal}
Michele Parrinello and Aneesur Rahman.
\newblock Crystal structure and pair potentials: A molecular-dynamics study.
\newblock \emph{Physical review letters}, 45\penalty0 (14):\penalty0 1196, 1980.

\bibitem[Lacroix et~al.(1998)Lacroix, Viguera, and Serrano]{lacroix1998elucidating}
Emmanuel Lacroix, Ana~Rosa Viguera, and Luis Serrano.
\newblock Elucidating the folding problem of $\alpha$-helices: local motifs, long-range electrostatics, ionic-strength dependence and prediction of nmr parameters.
\newblock \emph{Journal of molecular biology}, 284\penalty0 (1):\penalty0 173--191, 1998.

\bibitem[Jumper et~al.(2021)Jumper, Evans, Pritzel, Green, Figurnov, Ronneberger, Tunyasuvunakool, Bates, Žídek, Potapenko, Bridgland, Meyer, Kohl, Ballard, Cowie, Romera-Paredes, Nikolov, Jain, Adler, Back, Petersen, Reiman, Clancy, Zielinski, Steinegger, Pacholska, Berghammer, Bodenstein, Silver, Vinyals, Senior, Kavukcuoglu, Kohli, and Hassabis]{Jumper2021}
John Jumper, Richard Evans, Alexander Pritzel, Tim Green, Michael Figurnov, Olaf Ronneberger, Kathryn Tunyasuvunakool, Russ Bates, Augustin Žídek, Anna Potapenko, Alex Bridgland, Clemens Meyer, Simon A.~A. Kohl, Andrew~J. Ballard, Andrew Cowie, Bernardino Romera-Paredes, Stanislav Nikolov, Rishub Jain, Jonas Adler, Trevor Back, Stig Petersen, David Reiman, Ellen Clancy, Michal Zielinski, Martin Steinegger, Michalina Pacholska, Tamas Berghammer, Sebastian Bodenstein, David Silver, Oriol Vinyals, Andrew~W. Senior, Koray Kavukcuoglu, Pushmeet Kohli, and Demis Hassabis.
\newblock Highly accurate protein structure prediction with alphafold.
\newblock \emph{Nature}, 596:\penalty0 583--589, 8 2021.
\newblock \doi{10.1038/s41586-021-03819-2}.

\bibitem[Berman(2000)]{Berman2000}
H.~M. Berman.
\newblock The protein data bank.
\newblock \emph{Nucleic Acids Research}, 28:\penalty0 235--242, 1 2000.
\newblock \doi{10.1093/nar/28.1.235}.

\end{thebibliography}

\end{document}